\documentclass[fleqn,usenatbib]{mnras}

\usepackage{newtxtext,newtxmath}

\usepackage[T1]{fontenc}

\DeclareRobustCommand{\VAN}[3]{#2}
\let\VANthebibliography\thebibliography
\def\thebibliography{\DeclareRobustCommand{\VAN}[3]{##3}\VANthebibliography}

\usepackage{graphicx}	
\usepackage{amsmath}	

\usepackage{pdflscape}
\usepackage{siunitx}
\graphicspath{{./}{figures/}}
\usepackage{subfig}
\usepackage{multirow}
\usepackage{supertabular,booktabs}
\usepackage{arydshln}
\newcolumntype{L}[1]{>{\raggedright\arraybackslash}p{#1}}
\newcolumntype{C}[1]{>{\centering\arraybackslash}p{#1}}
\newcolumntype{R}[1]{>{\raggedleft\arraybackslash}p{#1}}

\newcommand{\subfigimg}[3][,]{%
  \setbox1=\hbox{\includegraphics[#1]{#3}}
  \leavevmode\rlap{\usebox1}
  \rlap{\hspace*{5pt}\raisebox{\dimexpr\ht1-1\baselineskip}{#2}}
  \phantom{\usebox1}
}

\title[iCOMs in outflows from G351.16+0.70]{Interstellar Complex Organic Molecules towards outflows from the G351.16+0.70 (NGC 6334 V) massive protostellar system}

\author[O. S. Rojas-García et al.]{
O. S. Rojas-García$^{1}$\thanks{E-mail: sergio.rojas@inaoep.mx},
A. I. Gómez-Ruiz$^{1}$,
A. Palau$^{2}$,
M. T. Orozco-Aguilera$^{1}$,
S. E. Kurtz$^{2}$, \newauthor
and M. Chavez Dagostino $^{1}$
\\
$^{1}$Astrophysics Department, Instituto Nacional de Astrofísica, Óptica y Electrónica, Luis E. Erro 1, Tonantzintla, Puebla, C.P. 72840, México\\
$^{2}$Instituto de Radioastronomía y Astrofísica, Universidad Nacional Autónoma de México, Antigua Carretera a Pátzcuaro \# 8701, \\ Ex-Hda. San José de la Huerta, Morelia, Michoacán, México C.P. 58089\\
}

\date{Accepted XXX. Received YYY; in original form ZZZ}

\pubyear{2015}

\begin{document}
\label{firstpage}
\pagerange{\pageref{firstpage}--\pageref{lastpage}}
\maketitle

\begin{abstract}
G351.16+0.70 is a relatively well-studied High Mass Star Forming Region with at least two main bipolar outflow structures originating from an OB embedded star and multiple IR cores. Using high-resolution and large bandwidth SMA observations, we studied its molecular content to probe the emission of iCOMs which could be related to the bipolar outflows or their jets. We analyzed the emission spectra in the 1mm band within 8 GHz bandwidth coverage, from 216.75 to 220.75 GHz, and from 228.75 to 232.75 GHz. Employing the LTE approximation using the XCLASS software, we identified 260 emission lines arising from iCOMs. The emission lines in the synthetic emission spectra could be explained by 11 iCOMs and 5 molecular isotopologues. Additionally, we analyzed the outstanding broad iCOM emission lines by using integrated and velocity field maps, searching for extended emission and velocity gradients related to molecular outflows. Ro-vibrational transitions of CH$_3$OH, C$_2$H$_3$CN, CH$_3$OCHO, CH$_3$COCH$_3$, and aGg'-(CH$_2$OH)$_2$ present evidence of extended emission that does not fit with spherical morphology and that follows the path of the low-velocity $^{13}$CO outflow. The multiple outflows in the system are revealed also by the CO (2--1) and SiO (5--4) emission, but in particular we have discovered a so-called extremely high velocity outflow ($|V_{Max}-V_{LSR}|\sim 60$ km s$^{-1}$). Additionally, we provide the full line catalog of iCOMs along the 8 GHz bandwidth produced by the main protostellar core.
\end{abstract}




\section{Introduction}
\label{sec:intro}

The search for molecules in massive outflows is partly motivated by the observed enhancement of certain molecules within shocked regions of well known low-mass outflows and their particular shock-induced chemistry. The prototype among these objects is the outflow driven by the low-mass protostar L1157-mm \citep{bachiller2001}, which has revealed a rich chemistry induced by shocks. Inhabiting this scenario, large molecules have also been observed, in particular those formed by at least six atoms, known as \textit{Interestellar Complex Organic Molecules }\citep[iCOMs;][]{vandishoek1998ARA&A..36..317V, leflo2017}. 
The presence of iCOMs tracing shocked regions 
suggests a relation between the shocks and iCOM formation or desorption \citep[][]{leflo2017}. The proposed two mechanisms to explain the abundances of iCOMs in several interstellar environments are dust grain chemistry and warm gas-phase reactions \citep{Garrod2006A&A...457..927G, Taquet_2012}.

A recent systematic study of the intermediate- to high-mass protostellar object IRAS 20126$+$4104 \citep[][]{palau2017}, indicates that iCOMs in this source may arise from the disc and dense/hot regions along the outflow. In fact, the abundances of some iCOMs show an enhancement at the outflow positions. iCOM enhancement has also been related to shocks in the accretion disc of HMSFR G328.2551-0.5321, a pre-Hot Core source \citep{Csengeri2019A&A,Csengeri2019A&A_1,Bouscasse2022A&A...662A..32B}. Recently, a sample of 11 massive protostars was investigated by means of single-dish mapping observations, revealing iCOM lines with faint broad profiles possibly related with large-scale low-velocity outflows \citep{Rojas_Garcia_2022}. In the present paper, we extend the study of iCOMs from massive outflows reporting the high-angular resolution observations of a prominent outflow system within the NGC6334 giant molecular cloud.  

This paper is organized as follows: section \ref{SEC:thesource} describes physical parameters obtained from the literature of G351.16+0.70. Section \ref{SEC:Observations} describes our observations and the ancillary surveys used in our analysis. In section \ref{SEC:xclass} we describe the LTE modeling used to catalog the observed iCOMs along with the strategy to look for their emission related to outflows. Section \ref{SEC:results} describes the integrated emission (moment 0) and velocity field (moment 1) maps for the more elongated iCOM emissions. Additionally, we compare our computed iCOM abundances with reported literature values, and spanning a range of star formation regimes, from low to high mass. Finally, section \ref{SEC:summaryandconclusions} presents a summary and the conclusions of this work.


\section{The source}\label{SEC:thesource}
G351.16+0.70 (G351, hereafter) is a relatively well-studied source, especially at infrared and centimeter wavelengths. It is related to the infrared source NGC 6334 V \citep{McBreen79,Kraemer99}, with a bipolar reflection nebula (likely related to a bipolar outflow driven by an unidentified, possibly massive, protostar \citep[e.g.][]{Chrysos94,Simpson09}), with a B-type zero-age main sequence and a massive embedded Young Stellar Object (YSO) \citep{Hashimoto2007}. It has been cataloged as an Extended Green Object (EGO) in the study of \citet{chen2013}. This Infrared (IR) Nebula had been associated with four IR cores \citep{Simon1985MNRAS.212P..21S,Kraemer1999ApJ...516..817K}, but after high-resolution polarimetric observations, \citet{Hashimoto2007} explain that these IR sources could be explained with only three YSOs and two massive outflows. 

This EGO has a bright core of $L\sim10^{5} L_\odot$ \citep{Loughran1986ApJ...303..629L} and is located at a distance of $1.3\pm0.3$ kpc \citep{Chibueze_2014}, with a $V_{LSR}$ of $-5.65 \pm 0.65$ kms$^{-1}$ \citep{Alvarez2004ApJS..155..123A}.  \citet{Fischer1982ApJ...258..165F} reported a high-velocity $^{12}$CO outflow coming from an embedded OB star; this was later reinforced by the combined analysis of mid-IR, far-IR, and radio-continuum observations of \citet{Harvey1984ApJ...280L..19H,Hashimoto2007}.   

Other evidence for the existence of a massive protostar includes a photo-dissociated region \citep[e.g.][]{Burton00}, CO high-velocity emission tracing a complicated structure, suggesting up to three outflows \citep{Juarez17}, and several previous studies suggesting that G351 harbors sources at several stages in the formation of massive stars \citep{Fischer1982ApJ...258..165F,Harvey1984ApJ...280L..19H,Kraemer1999ApJ...516..817K,Hashimoto2007}.


\section{Observations, data reduction, and ancillary data}\label{SEC:Observations}
In this study, we used observations of two published surveys, the {\it Spitzer} Galactic Legacy Infrared Midplane Survey Extraordinaire (GLIMPSE\footnote{\href{https://irsa.ipac.caltech.edu/data/SPITZER/docs/irac/}{https://irsa.ipac.caltech.edu/data/SPITZER/docs/irac/}}) and the APEX Telescope Large Area Survey of the Galaxy (ATLASGAL\footnote{\href{http://atlasgal.mpifr-bonn.mpg.de/cgi-bin/ATLASGAL\_DATABASE.cgi}{http://atlasgal.mpifr-bonn.mpg.de/cgi-bin/ATLASGAL\_DATABASE.cgi}}), along with our own observations with the Submillimeter Array (SMA\footnote{\href{https://lweb.cfa.harvard.edu/sma/}{https://lweb.cfa.harvard.edu/sma/}}).

\subsection{Submillimeter Array}
Our observation was carried out on May 11 of 2011, pointing toward G351 (R.A 17:19:57.70, Dec -35:57:50.0 J2000) with the SMA, a facility of the Smithsonian Astrophysical Observatory, on Mauna Kea, Hawaii.

We used the SMA compact configuration, with baselines ranging from 16.4 m to 77.0 m. The 230 GHz receiver was configured in the 8 GHz Dual Rx mode, and tuned to 217.105 GHz for a simultaneous SiO (5-4) and CO (2-1) observation. The bandwidth coverage goes from 216.75 to 220.75 in the lower sideband (LSB) and from 228.75 to 232.75 GHz in the upper sideband (USB), and has a spectral resolution of $\sim 812$ KHz  ($\sim$1.1 kms$^{-1}$), with a mean rms of 90 mJy beam$^{-1}$channel$^{-1}$. At these frequencies, the primary beam size ($\theta_{HPBW}$) of each 6 m diameter antenna is $\sim$54$''$. The data is currently available for public use in the SMA archive\footnote{\href{https://lweb.cfa.harvard.edu/cgi-bin/sma/smaarch.pl}{https://lweb.cfa.harvard.edu/cgi-bin/sma/smaarch.pl}}. 
The visibility data were calibrated using the MIR software package, which was originally developed for the Owens Valley Radio Observatory \citep{Sco93} and adapted for the SMA\footnote{\href{https://lweb.cfa.harvard.edu/~cqi/mircook.html}{https://lweb.cfa.harvard.edu/$\sim$cqi/mircook.html}}. The absolute flux density scale was determined from observations of Neptune. A pair of nearby compact radio sources, 1626-298 and 1802-396, were used to calibrate the relative amplitude and phase. We used 3C279 to calibrate the bandpass.

The image map conversion from \textit{UV} visibility data was conducted using the \textit{Robust weighting} method, through \textit{INVERT} task in the Miriad software. This weighting is a compromise between Natural and Uniform weightings and is commonly used to image faint extended emission, therefore applicable to the outflow nature of our source, to avoid the loss of extended emission and also get good suppression of the side lobe contribution. The functional parameters of this weighting were defined as follows: Tapering of the data (FWHM)= 1$''$, Suppression area (SUP)= full coverage of the map, Robust parameter (Robust)= 2, Visibility weight= 1/$\sigma$. After this inversion, we obtained a map with an angular resolution of $\theta_{synth}=5.3'' \times 3.6''$ with a position angle of 74.1$^{\circ}$.

During the image deconvolution process, we smoothed the spectral resolution to 2.2 kms$^{-1}$ (two channels wide), to reduce our RMS level to 68 mJy beam$^{-1}$channel$^{-1}$. We take this conservative value considering to keep good spectral resolution to avoid blending in the emission lines but improving the signal to noise ratio, as we expected that wing profile features were weak. To convert the flux density (Jy) to $T_{mb}$, we used the conversion factor obtained from the \textit{IMSTAT} miriad task for each datacube ($\sim 1.25 K/(Jy/beam)$).

\subsection{Continuum observations at 870$\mu$m}

Dust continuum maps at 870$\mu$m were obtained from the ATLASGAL survey (Schuller et al., 2009), through The ATLASGAL database Server, that provides FITS format maps with a size of 5x5 arcmin centered on the peak continuum source. The details of these survey observations are presented in \cite{schuller2009}.

\subsection{Spitzer IRAC}
Images of 3.6$\mu$m, 4.5$\mu$m and 8.0$\mu$m IRAC bands, in the framework of the GLIMPSE survey, were retrieved from the NASA/IPAC Infrared science data archive. We used the post-basic calibrated data (BCD) images for our study. The mean FWHM of the point spread functions are 1.66$''$, 1.72$''$, and 1.98$''$ for bands 1, 2, and 4, respectively.

\section{XCLASS LTE Modeling Implementation}\label{SEC:xclass}
For our spectral line analysis we take the emission spectrum of a single synthesized beam pointing ($\theta_{synth}=5.3'' \times 3.6''$)  at the brightest core position (Ra 17h 19m 57.373s; Dec -35d 57m 52.00s), and, from the lower to the upper sideband we searched for line emission included in the XCLASS \citep{xclass} molecular database\footnote{\href{https://xclass.astro.uni-koeln.de/}{https://xclass.astro.uni-koeln.de/}} (CDMS/VAMDC) by means of their \textit{GetTransitions()} task. 
We listed all molecules within a $V_{LSR}$ maximum offset of $\pm$2 kms$^{-1}$ and a $T_{Rot}$ from 10 to 950 K and visually selected molecular candidates considering their frequency coincidence, upper state energy, and line velocity offset.  

To reinforce our line catalog, we implemented an LTE synthetic modeling employing the XCLASS 
task \textit{myXCLASSFit()}. The grid was computed with T$_{Rot}$ (rotational temperature in K), $N_{tot}$ (total column density in cm$^{-2}$), $V_{width}$ (velocity line width in km s$^{-1}$) and $V_{off}$ (velocity offset in kms$^{-1}$) as free parameters.   

The T$_{Rot}$ were limited from 10 to 900 K, the $V_{width}$ went from 1 to 7 kms$^{-1}$ for the narrow emission lines and from 5 to 15 kms$^{-1}$ for the wider ones. The $N_{tot}$ values varies with the species, with lower and higher limits of \num{1E13} and \num{1E17} cm$^{-2}$, according to their extreme reported values \citep{Csengeri2019A&A, DeSimone2020A&A...640A..75D}.  

To set the velocity offset, we take the velocity of the core tracer CH$_3$OH. As we see an asymmetrical distribution of the peak of its lines we take the outer values of $V_{off}=  -3.8$ to 2.2 kms$^{-1}$ away from the $V_{LSR}$. 

Finally, the source size parameter was fixed to 5$''$, as this is the spatial extent covered by the $\theta_{synth}$ of our data and the emission fills the beam, therefore we could assume a beam filling factor $\Omega_S/\Omega_A=1$. 

The resulting physical parameters were found computing an \textit{algorithm chain}, consisting of a particle swarm optimization method named \textit{Bees} \citep{Bees2005}, linked to the \textit{Levenberg-Marquardt} method \citep{Marquardt1963}, an optimization of \textit{Gauss-Newton} and \textit{Descendent Gradient} methods \citep{xclass}. 

\begin{figure*}
	\centering  
	\includegraphics[width=\linewidth]{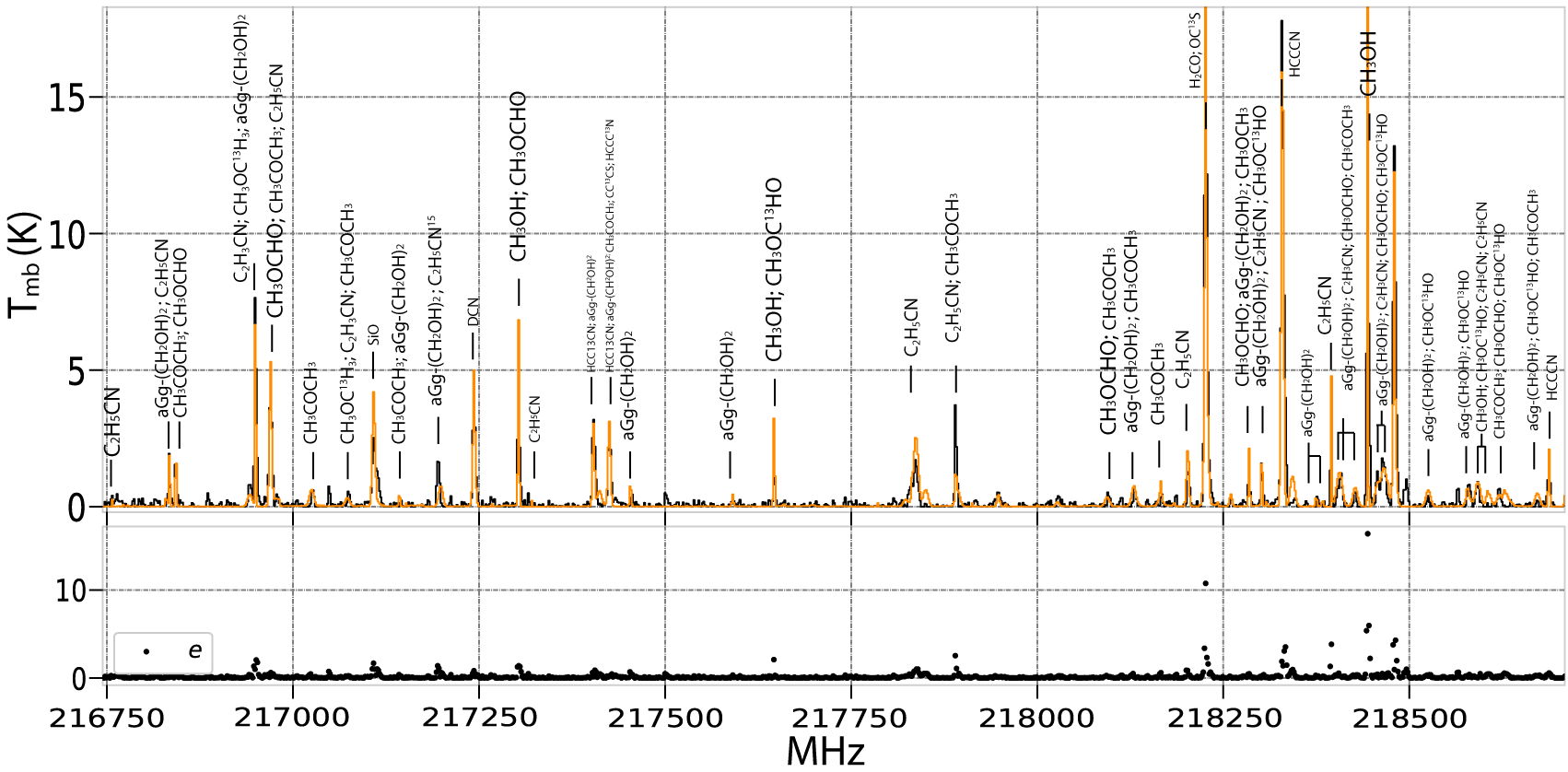}
	\caption{G351 spectrum with identified iCOMs (labeled) towards the brightest core KDJ4 and their corresponding fit modeling using XCLASS. The \textit{Black} solid line shows the observed spectrum, the \textit{Orange} solid line shows the modeled spectrum obtained using XCLASS. In the lower subplot, we show the residuals ($e = I_{model}-I_{obs}$).}
	\label{216-218}
\end{figure*}

The Goodness of fit between our models against the observed spectra was tested by a classic $\chi ^{2}$ test. Using the number of channels in each bandwidth as our degree of freedom, we computed a Critical $\chi^{2}$ corresponding to a significance level of a p-value = 0.05 and set it as our goal.  All of our models have successfully passed this threshold, strengthening our line classification. Illustratively, we show a plot of our computed modeled spectra over-plotted on the observed one in Figure \ref{216-218}. In the lower part of the plot we show the residuals ($e = I_{model}-I_{obs}$). 

After the species identification, we looked for broad iCOM emission and computed the moment 0 and moment 1 maps 
in the stronger emission lines in order to find outflow tracers considering their elongated spatial extent. 

\section{Results}\label{SEC:results}
\subsection{The structure of G351: envelope, cores and outflows}\label{subsec:Structure_g351}
\subsubsection{Mid-Infrared emission}
\begin{figure}
\centering
\includegraphics[angle=0,width=9cm]{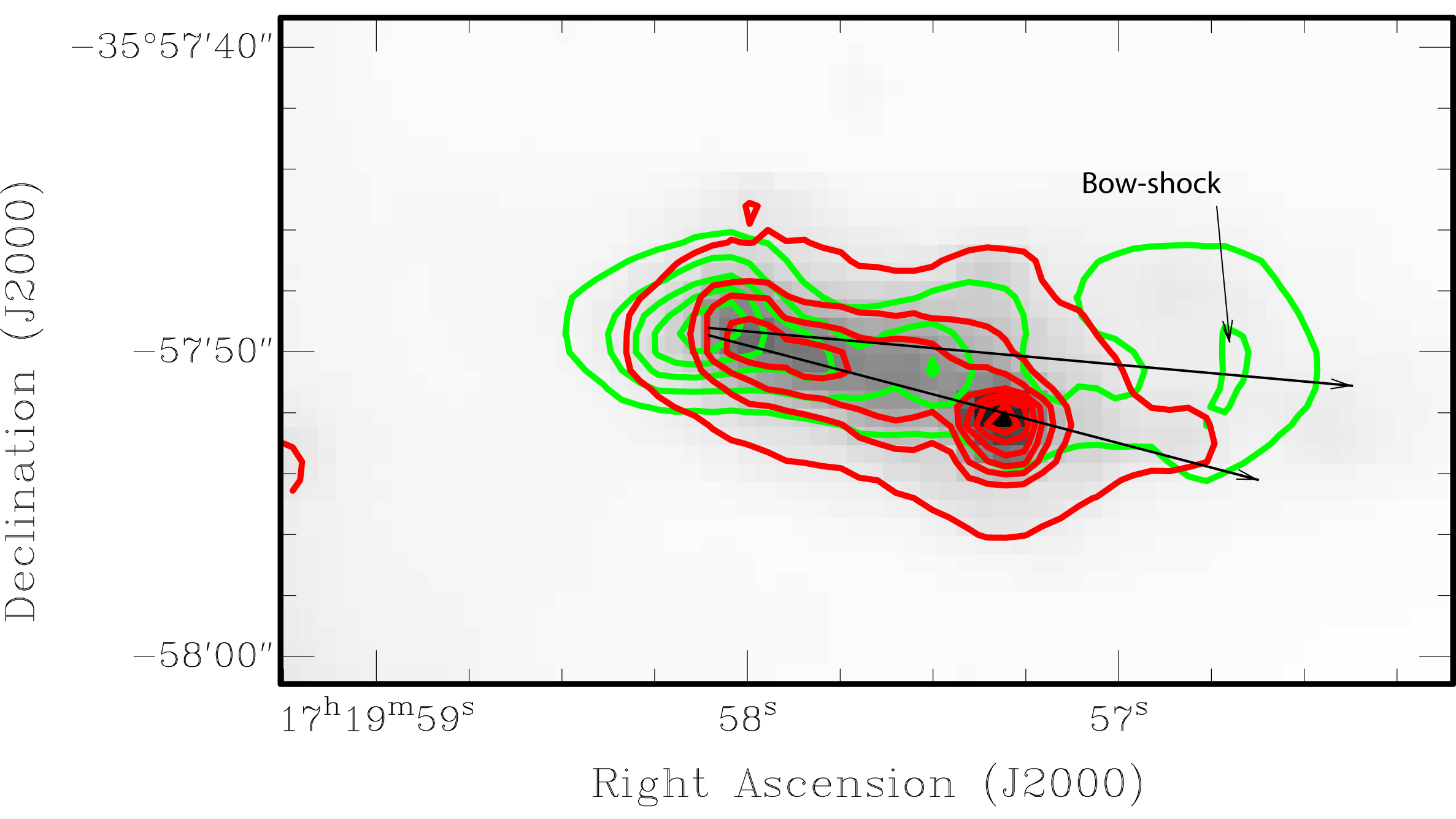}
\caption{The 4.5 $\mu$m (green contours) and 8.0$\mu$m emission (grey scale and red contours) in G351. Contour spacing and first contour is 10\% of the flux peak (2135 and 4653 MJy sr$^{-1}$, for 4.5 $\mu$m and 8.0 $\mu$m, respectively). Arrows show the axes of the two possible ejection events. Also indicated by an arrow is the bow-shock structure.}
\label{G351-chans-2-4}
\end{figure}

Considering the GLIMPSE three color images presented by \citet[][blue: 3.6$\mu$m, green: 4.5$\mu$m, red: 8.0$\mu$m]{Cyganowski2008} we infer a partially bipolar structure with the emission showing a {\it green excess}, oriented in the east-west direction with a very clear bow shock structure in the western lobe (Figure \ref{G351-chans-2-4}). This structure is located inside a region that is dark at 8$\mu$m and was previously shown in \citet{Juarez17}. 

Additionally, a collimated structure is seen in the 8$\mu$m image, but pointing towards a slightly different direction to the 4.5$\mu$m bow-shock (see Fig. \ref{G351-chans-2-4}). This may indicate that either the east-west outflow is precessing or the collimated 8$\mu$m structure is tracing an outflow different to the one that produced the 4.5$\mu$m bow-shock. Also, at the tip of this collimated structure there is a compact 8$\mu$m emission peak, which coincides with a dust continuum core (see below).  

\subsubsection{Millimeter dust continuum: envelope and cores}
\begin{figure}
\begin{center}
\includegraphics[width=\linewidth]{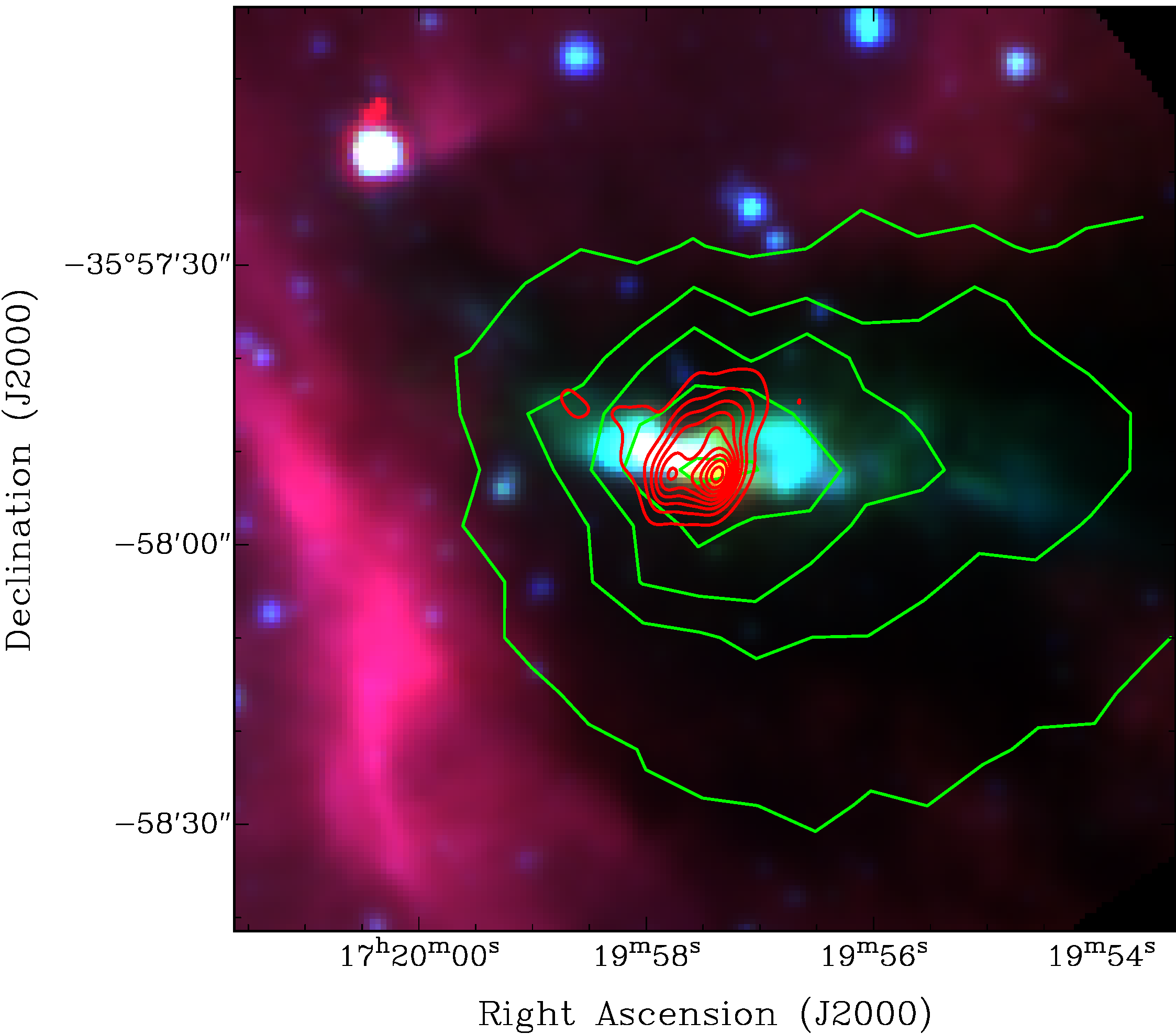}
\caption{Map of the continumm emision towards G351. The red contours show the 1.4 millimeter continuum emission, whereas the green contours show the 870 $\mu$m emission. In both contour maps the first contour starts at 20\% of the peak emission with a contour spacing of 10\%. The background shows the 3 color mid-IR image from Spitzer. }
\label{continuum}
\end{center}
\end{figure}

The continuum millimeter emission from G351 is presented in Fig \ref{continuum} over the 3 color composite GLIMPSE image. The 1.4 millimeter continuum shows at least two peaks that are separated by a few arc-sec, and agrees with the two confirmed cores by \citet{Hashimoto2007,Juarez17}. Its extended emission is seen mainly toward the north of these continuum peaks. The 1.4 mm emission coincides with peak contour at 870 $\mu$m and both peaks are very close to the brightest core KDJ4. However, the extended emission at 870$\mu$m is not seen in the 1.4 mm continuum image, but this may be because this emission is outside the primary beam along with the change in sensitivity for these observations. In Table \ref{cont-param} the results of Gaussian fits to the two peaks at 1.4 mm, component 1 and 2, are presented.  

\begin{table}
\caption{Parameters of the 1.4 mm continuum two main peaks obtained using a Gaussian fit by \textit{imfit} task in miriad.}
\centering
\begin{tabular}{l c c}
\hline
\hline
&Component 1 & Component 2\\
\hline
Peak intensity (Jy/beam) & 0.85$\pm$0.08 & 0.56$\pm$0.04\\
Total integrated flux (Jy) & 1.77$\pm$0.18  & 1.45$\pm$0.11\\
R.A. (J2000) & $17^h19^m57.41^s$ & $17^h19^m57.73^s$\\
DEC (J2000)     & $-35^{\circ}57'52.3''$ & $-35^{\circ}57'52.27$ \\
Deconvolved Major  axis (arcsec)  &    4.1$\pm$0.97 & 4.9$\pm$1.15\\
Deconvolved Minor axis (arcsec)  &  3.4$\pm$0.53  & 4.3$\pm$0.81\\
Deconvolved Position angle (degrees) & -46.3 &  -7.2 \\
\hline
\end{tabular}
\label{cont-param}
\end{table}

\subsubsection{High-Velocity and Extremely High-Velocity emission in G351}
\begin{figure}
	\centering  
\includegraphics[width=\linewidth]{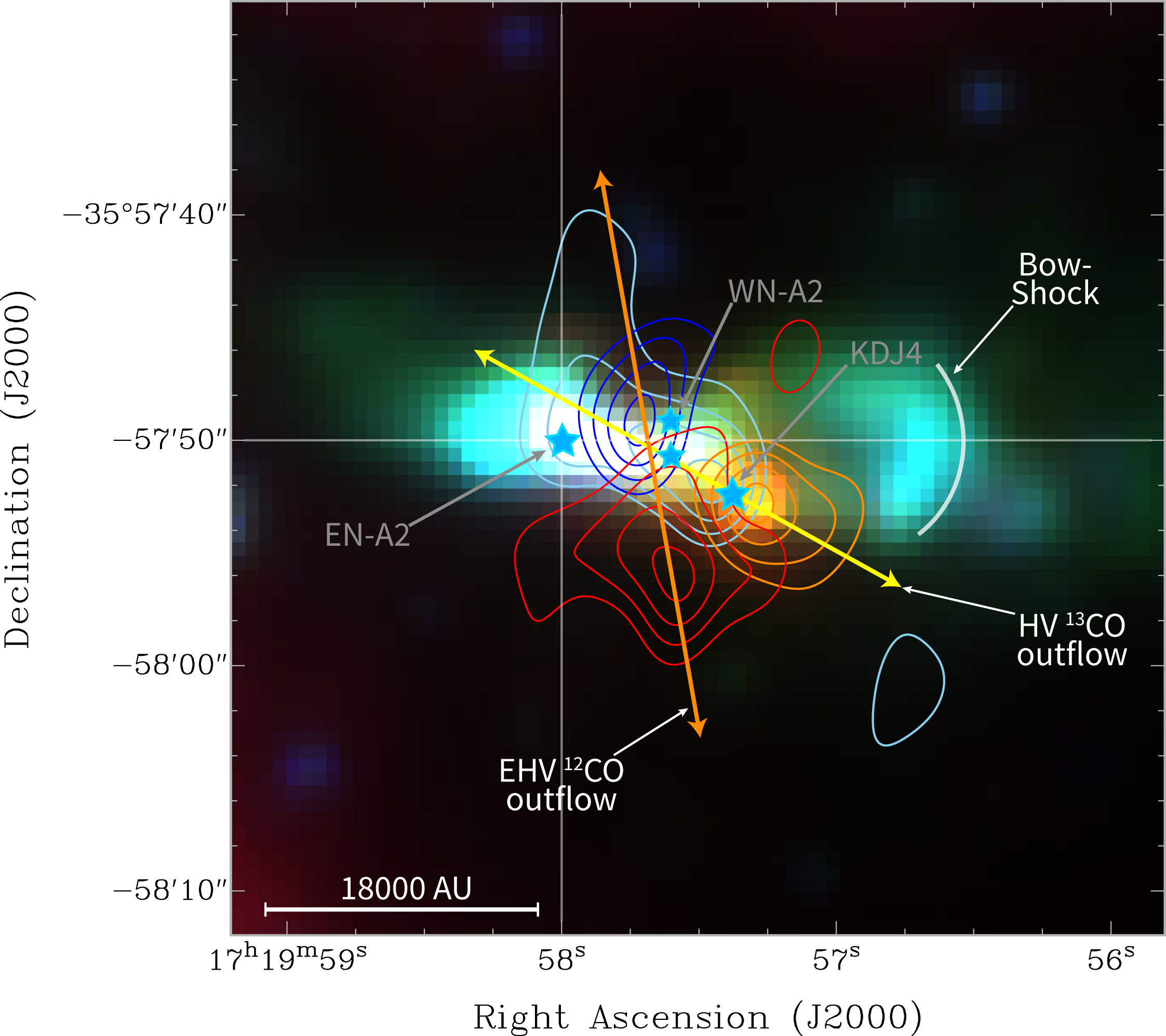}
	\caption{G351 Integrated map of the two bipolar outflows traced by $^{12}$CO (EHV) and $^{13}$CO (HV). The star markers show the near- to far-IR confirmed cores by polarimetric observations \citep{Kraemer1999ApJ...516..817K, Hashimoto2007}. The background color image is a three band composite for 3.6, 4.5, and 8.0$\mu m$, coded as blue, green, and red, respectively \citep[][]{Cyganowski2008}. The blue and red contours show the blue- and red-shifted Extremely High Velocity outflow ($|V_{Max}-V_{LSR}|\sim 60$ kms$^{-1}$) traced by $^{12}$CO. Correspondingly, the sky blue and orange contours display the blue- and red-shifted emission for High-Velocity outflow traced by $^{13}$CO ($|V_{Max}-V_{LSR}|\sim 20$ kms$^{-1}$). Contour spacing is 3$\sigma$ and first contour is 6$\sigma$.}
	\label{Outflow_scheme}
\end{figure}

We detected CO (2-1) emission going from $\sim$$-$65 to +27 km s$^{-1}$, which therefore reaches the classification of an Extremely High Velocity outflow \citep[EHV: $|V_{Max}-V_{LSR}|\sim 65$ km s$^{-1}$,][]{tafa2010}.  Figure \ref{Outflow_scheme} shows the CO (2-1) emission integrated over the high-velocity wings (blue: -65.0 to -55.0 km s$^{-1}$, red: +17.0 to +27.0 km s$^{-1}$), overlaid on the three color GLIMPSE image. The red- and blue-shifted wings are oriented in a north-south direction, with the outflow center closer to continuum component 2 than to component 1, i.e. located  $\sim$6$''$ to north-east from the core KDJ4 \citep{Kraemer1999ApJ...516..817K} and coinciding with the near- to far-IR core WN-A2. The spatial extent of this emission is $\sim$\num{22.6E3} AU. This size and an assumed constant velocity of 60 km s$^{-1}$ suggets a dynamical age of $t_{dyn}=1800 yr$.

 We also detected emission at High Velocities employing the high-density tracer $^{13}$CO (2-1) \citep[HV: $|V_{Max}-V_{LSR}|\sim 35$ km s$^{-1}$][]{tafa2010}. This outflow seems to emerge from the core KDJ4 \citep{Kraemer1999ApJ...516..817K} and nearly follows the east-west path presented by \citet{Juarez17}, as it shows a P.A. of  155.4$^\circ$, and is spread away for $\sim$ \num{23.2E3} AU with an observed maximum velocity of $|V_{Max}-V_{LSR}|\sim20$ kms$^{-1}$. Taking these values and following the same assumptions as before, this object has a dynamical age of $t_{dyn}=5500 yr$. Figure \ref{Outflow_scheme} shows a labeled scenario of G351 including both the CO (2-1) EHV and the $^{13}$CO (2-1) HV outflow.

\begin{figure}
\begin{center}
\includegraphics[width=\linewidth]{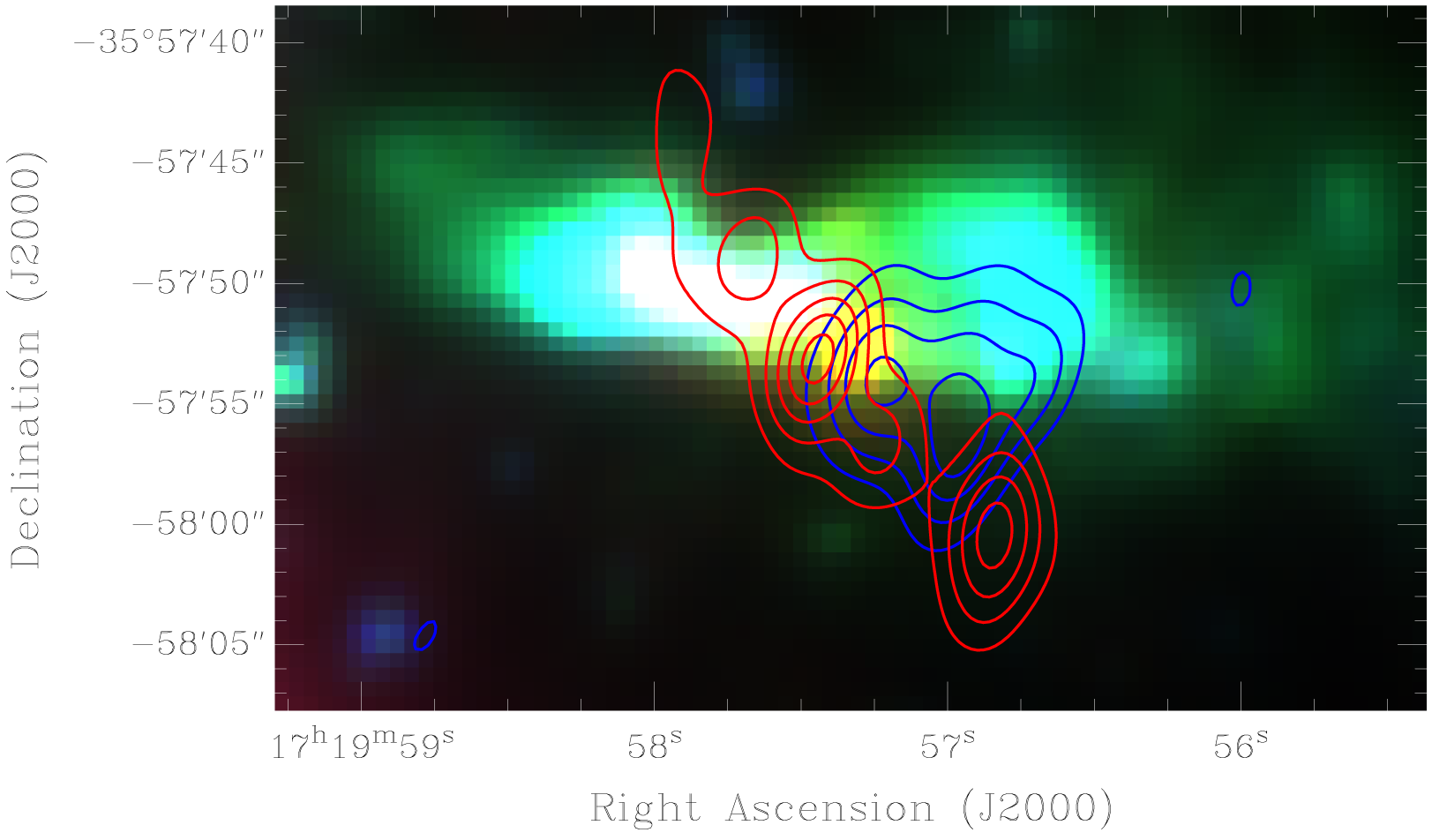}
\caption{G351 Integrated map of the outflow wings traced by SiO (5-4) emission. Contour spacing and first contour is 4$\sigma$ ($\sigma =$0.199 and 0.174 Jy beam, SiO blue-shifted and red-shifted, respectively). The background color image is three bands composite for 3.6, 4.5, and 8.0$\mu m$, coded as blue, green, and red, respectively \citep[][]{Cyganowski2008}. The blue and red contours shows the blue- and red-shifted HV outflow traced by SiO ($|V_{Max}-V_{LSR}|\sim 20$ kms$^{-1}$).
}
\label{G351_16_CO-SiO}
\end{center}
\end{figure}

The SiO (5-4) emission was detected in a narrower velocity range than the CO (2-1) emission but still having a blue-shifted component of HV (blue: -23.2 to -18.0 km s$^{-1}$, red:-2.4 to +0.2 km s$^{-1}$). The velocity integrated emission map is shown in Fig \ref{G351_16_CO-SiO}; it shows blue shifted emission that agrees with the P.A. of the HV outflow traced by $^{13}$CO. Nevertheless, the red-shifted emission shows a completely different orientation than the other red-shifted emissions, but seems to follow the blue shifted HV of $^{13}$CO. Thus, the interpretation of the SiO emission is not straightforward, but it is possible that it traces an additional outflow. In fact, the red- and blue-shifted SiO emission has a similar sky position with the southern region of the converging flow traced by H$^{13}$CO$^{+}$ \citep{Juarez17}, but in this case at lower velocities (blue: -9.2 to -7.1 km s$^{-1}$, red:-5.0 to -2.9 km s$^{-1}$).  

Previous studies have shown differences between the structures traced by CO and SiO. Examples of similar cases include the Orion-S region \citep{Zapata06} and  IRAS 05358$+$3543 \citep{Beuther02}, which were also observed with interferometers. Apart from observational effects such as spatial filtering, the difference between the SiO and CO outflow emission can be attributed to shock chemistry effects \citep{Gusdorf08b}. For example, from shock models in low-mass systems it is known that the shock velocity required to form SiO may not be reached in all outflows \citep{Gusdorf08b}. In addition, observational evidence in low-mass outflows shows that evolved outflows do not present prominent SiO emission \citep{Bachiller96}. However, it is important to point out that models to explain the SiO emission in massive systems are scarce, and therefore it is not obvious whether a similar explanation as in low-mass systems can be applied to massive outflows \citep{Leurini13}.   

\subsubsection{Low-Velocity emission in G351}

In our interferometric observations, emission at Low Velocity in CO (2-1) was not detected \citep[LV: $|V_{Max}-V_{LSR}|\sim 10$ km s$^{-1}$][]{tafa2010}, but this is possibly due to a combination of spatial filtering and contamination from the diffuse emission that hides the CO (2-1) low velocity emission.
In general, we have noted missing structures going southwest from KDJ4, especially  in the extended emission of the simplest molecules such as SiO, CO, and their isotopologues (see below).

It is relevant to note that the outflow emission revealed by the SMA maps of the CO (3-2) line presented by \citet{Juarez17} actually trace the LV component of the east-west and north-south outflows, but not the high-velocity component of the north-south outflow that dominates the outflow emission in our CO (2-1) wings-map. The fact that the CO (3-2) line reveals the east-west outflow indicates that, similar to other \textit{High-J} CO lines, the low-velocity outflow emission is not severely affected by the ambient emission in these transitions. The non-detection of the high-velocity emission of the north-south outflow in the CO (3-2) line may be due to a combined effect of sensitivity and excitation.

\subsection{XCLASS identification catalogue}
The iCOM identification was conducted in the spectrum of the brightest core, historically named KDJ4 \citep{Kraemer1999ApJ...516..817K,Hashimoto2007}. From our analysis, we identified 11 iCOM main isotopologues, five secondary isotopologues and two vibrationally excited states\footnote{For CH$_3$OH and CH$_3$OCHO} in G351, most of them having already been detected in multiple SFRs from low to high mass regimes \citep{Ospina-Zamudio2018A&A...618A.145O}. The chemical iCOMs species observed in this \textit{Core} are listed in Table \ref{iCOMs_Full_Table}. The simpler molecules' emission in G351 will be discussed in a separate paper.  
\begin{table}
	\centering
	\caption{Detected iCOM species sorted by complexity.}
	\begin{tabular}{ccccccccccccccccc}
		\toprule
         Main species    & Isotopologue/Isotopomer \\
		\midrule
		CH$_3$COCH$_3$  &  \\
		& aGg'-(CH$_2$OH)$_2$  \\
		CH$_3$OCH$_3$  &  \\
		C$_2$H$_5$OH   &  \\
		C$_2$H$_5$CN   & C$_2$H$_5$C$^{15}$N, C$_2$H$_5$$^{13}$CN \\
		CH$_3$OCHO & CH$_3$O$^{13}$CHO  \\

		C$_2$H$_3$CN &   \\
		NH$_2$CHO   & \\
		CH$_3$CHO   &      \\
		CH$_3$OH  &  \\
		CH$_3$CN  &  CH$_3$$^{13}$CN, $^{13}$CH$_3$CN  \\
		\bottomrule                                                  
	\end{tabular}\label{iCOMs_Full_Table}
\end{table}

After the modeling, we were able to discriminate some low-intensity lines from several line candidates, so that the complete chemical transition list increased. We have observed about 175 emission lines with intensities over $T_{MB} > 0.5$ K within our 8 GHz bandwidth. The complete list of detected transitions for all the identified iCOMs is reported in Table \ref{iCOMs_catalogue_Example}. 
\begin{table}
	\centering
	\caption{Modeled lines with T$_{mb}>$ 0.8. by means of the XCLASS identification model. The columns shows: (1) The molecular name and the vibrational excited state, (2) the rest frequency, (3) the spontaneous emission coefficient A$_{ij}$ and (4) the Upper State degeneracy. In this table we only present an example of the format of detected transitions, the complete list of the 260 high intensity transitions is reported in appendix \ref{APx_Full_List}.}
		\footnotesize 
	\begin{tabular}{ccccccccccccccccc}
		\toprule
\multirow{2}{2.2cm}{iCOM species}    & Frequency & A$_{ij}$ &  g$_{up}$\\
  &  [GHz]& [s$^{-1}$]&  \\
		\midrule
C2H5CN;v=0;			&	216.8281      &	\num{6.9446E-07}	&	159 \\
CH3OCHO;v=0;		&	216.8302      &	\num{1.4796E-04}	&	74 \\
CH3OCHO;v=0;		&	216.8389      &	\num{1.4800E-04}	&	74 \\
CH3OH;v=0;			&	216.9456      &	\num{1.2135E-05}	&	44 \\
aGg'-(CH2OH)2;v=0;	&	216.9466      &	\num{5.1108E-06}	&	525 \\
CH3OCHO;v=0;		&	216.9630      &	\num{2.4448E-05}	&	82 \\
CH3OCHO;v=0;		&	216.9642      &	\num{2.4436E-05}	&	82 \\
CH3OCHO;v=0;		&	216.9648      &	\num{1.5313E-04}	&	82 \\
CH3OCHO;v=0;		&	216.9659      &	\num{1.5315E-04}	&	82 \\
CH3OCHO;v=0;		&	216.9662      &	\num{1.5313E-04}	&	82 \\

		\bottomrule                    
	\end{tabular}\label{iCOMs_catalogue_Example}
\end{table}

\subsection{iCOMs along the outflows}
We have observed that although most of the iCOMs peak at KDJ4 core, several of them have emission spread along the High-Velocity outflow direction, but also tracing Low Velocity (LV) gas $|V_{Max}-V_{LSR}| < 10$ kms$^{-1}$.

As the velocity space covered by iCOMs and the simpler molecules shows a discrepancy of tens of kms$^{-1}$, we tried to establish the association of iCOMs with the outflows, looking for coincidence in the sky-plane locations. For this, we analyzed the moment 0 and moment 1 maps of the stronger iCOM emission, finding 10 iCOM species presenting an elongated extension (with a major axis longer than $\sim10''$), and five transitions tracing the bow shock \textit{EGO's} structure (Figure \ref{Outflow_scheme}). Since this emission departs from the spherical geometry expected for a compact core, and because the extended emission direction overlaps with the bipolar outflows traced by simpler molecules, we concluded that the iCOM emission is coming from the outflow itself. 

The iCOMs found to trace the outflow are CH$_3$OH, CH$_3$OCHO, C$_2$H$_3$CN and C$_2$H$_5$CN. In some cases, these lines are blended with CH$_3$OCH$_3$, CH$_3$COCH$_3$ and aGg-(CH$_2$OH)$_2$. We recall that these iCOMs' emission were visually inspected to avoid the emission lines blended with simpler molecules, i.e. the analyzed emission lines, even when blended, are only produced by iCOM species. It is also important to note that, similar to the diatomic molecules, spatial filtering seems to affect the low-velocity emission of more extended iCOMs, such as Methanol and Vinyl cyanide. 
The details of the molecular emission are described in Table \ref{Extended_iCOM_Table}.

\begin{table*}
	\centering
	\caption{List of iCOMs detected in the outer regions of the Core, their corresponding transition frequency is shown in the second column, whereas the third column shows their corresponding transition.} 
	\begin{tabular}{ccccccccccccccccc}
		\toprule
         Species    & Frequency [MHz]&  Transition\\
		\midrule
		CH$_3$OH                                  &216945.6000 & v12 = 0; J = 5 - 4 \\
		aGg'-(CH$_2$OH)$_2$	                      &216946.5593 & vibInv = s-a; J = 37$_{8}$ - 36$_{9}$  \\ \cdashline{1-2}[.3pt/3pt] 
		CH$_3$OCHO                                &218280.9000 & v18 = 0; J = 17$_{3,14}$ - 16$_{3,13}$\\
		CH$_3$COCH$_3$                            &218281.3692 & J = 48$_{22,26}$ - 48$_{21,27}$ (AA)\\ 
		aGg'-(CH$_2$OH)$_2$                       &218283.0917 & vibInv = s-a; J = 9$_{7,3}$ - 8$_{6,3}$\\
				\cdashline{1-2}[.3pt/3pt] 
		C$_2$H$_3$CN 	                          &218421.8013 & J = 23$_{8,16}$ - 22$_{8,15}$\\
				\cdashline{1-2}[.3pt/3pt] 
		CH$_3$OH	                              &220078.5610 & v12 = 0; J = 8$_{0}$ - 7$_{1}$ \\
				\cdashline{1-2}[.3pt/3pt] 
		CH$_3$CN	                              &220641.0839 & J = 12$_{5}$ - 11$_{5}$\\
				\cdashline{1-2}[.3pt/3pt] 
	    CH$_3$OCHO	                              &229405.0210 & v18 = 0; J = 18$_{3,15}$ - 17$_{3,14}$\\
	    		\cdashline{1-2}[.3pt/3pt] 
		CH$_3$OH	                              &229758.7560 & v12 = 0; J = 8$_{1}$ - 7$_{0}$\\
				\cdashline{1-2}[.3pt/3pt] 
		CH$_3$OH                                  &230027.0470 & v12 = 0; J = 3$_{2}$ - 4$_{1}$\\
		CH$_3$COCH$_3$                            &230028.3815 & J = 19$_{16,4}$ - 19$_{13,7}$\\
				\cdashline{1-2}[.3pt/3pt] 
		C$_2$H$_5$CN                              &231313.2412 & J = 24$_{2,23}$ - 23$_{1,22}$\\
		CH$_3$OCHO	                              &231315.0300 & v18 = 0; J = 29$_{4,27}$ - 29$_{3,26}$\\
				\cdashline{1-2}[.3pt/3pt] 
		C$_2$H$_5$CN                              &231990.4098 & J = 27$_{0,27}$ - 26$_{0,26}$\\
		CH$_3$OCHO	                              &231985.3570 & v18 = 0; J = 20$_{9,12}$ - 20$_{8,13}$ (A)\\
		\bottomrule                    
	\end{tabular}\label{Extended_iCOM_Table}
\end{table*}

\subsubsection{Methanol (CH$_3$OH)} 
\begin{figure*}
  \centering
  \begin{tabular}{@{}p{0.33\linewidth}@{\quad}p{0.33\linewidth}@{}p{0.33\linewidth}}
    \subfigimg[width=\linewidth]{a)}{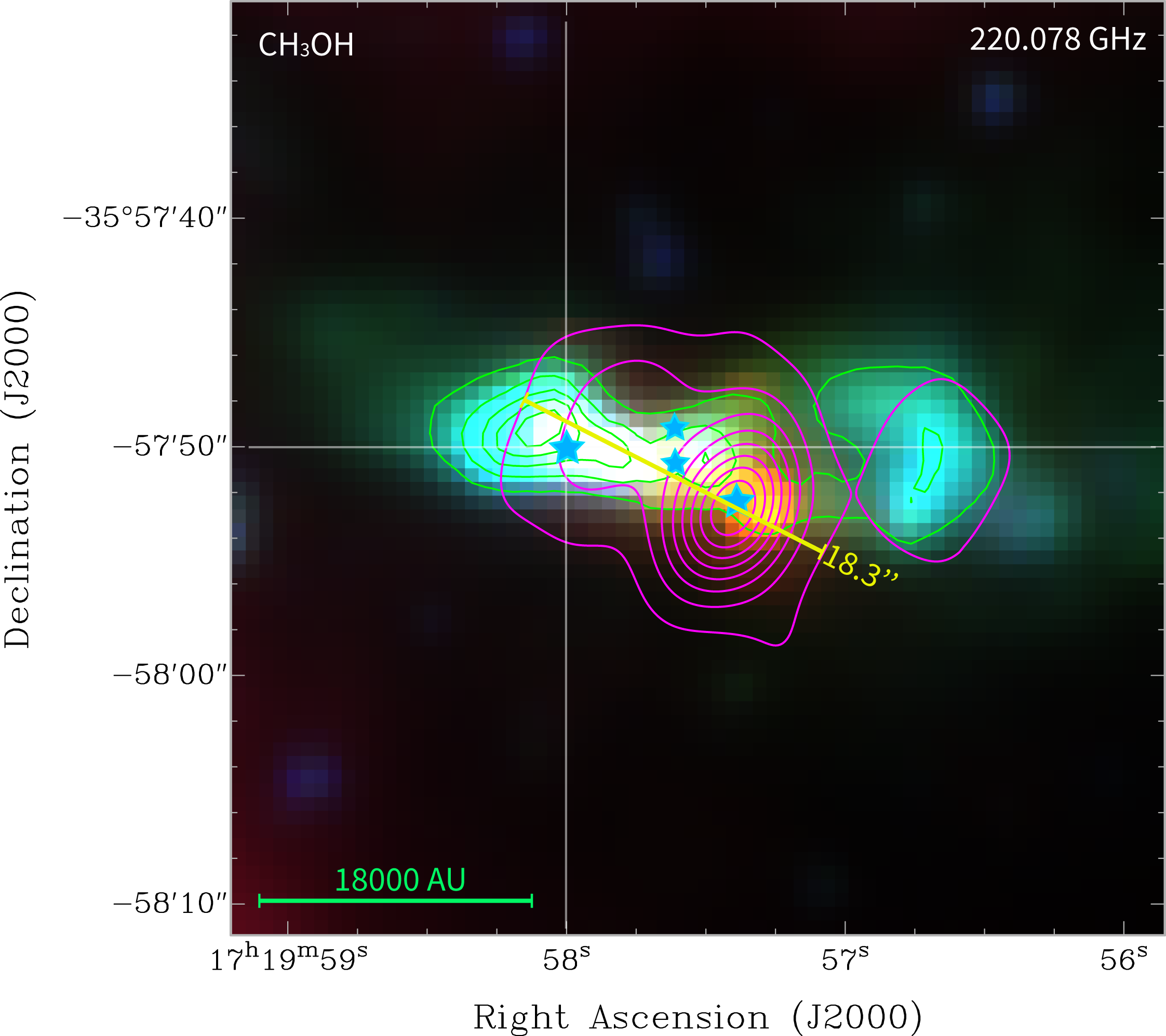} &
    \subfigimg[width=\linewidth]{b)}{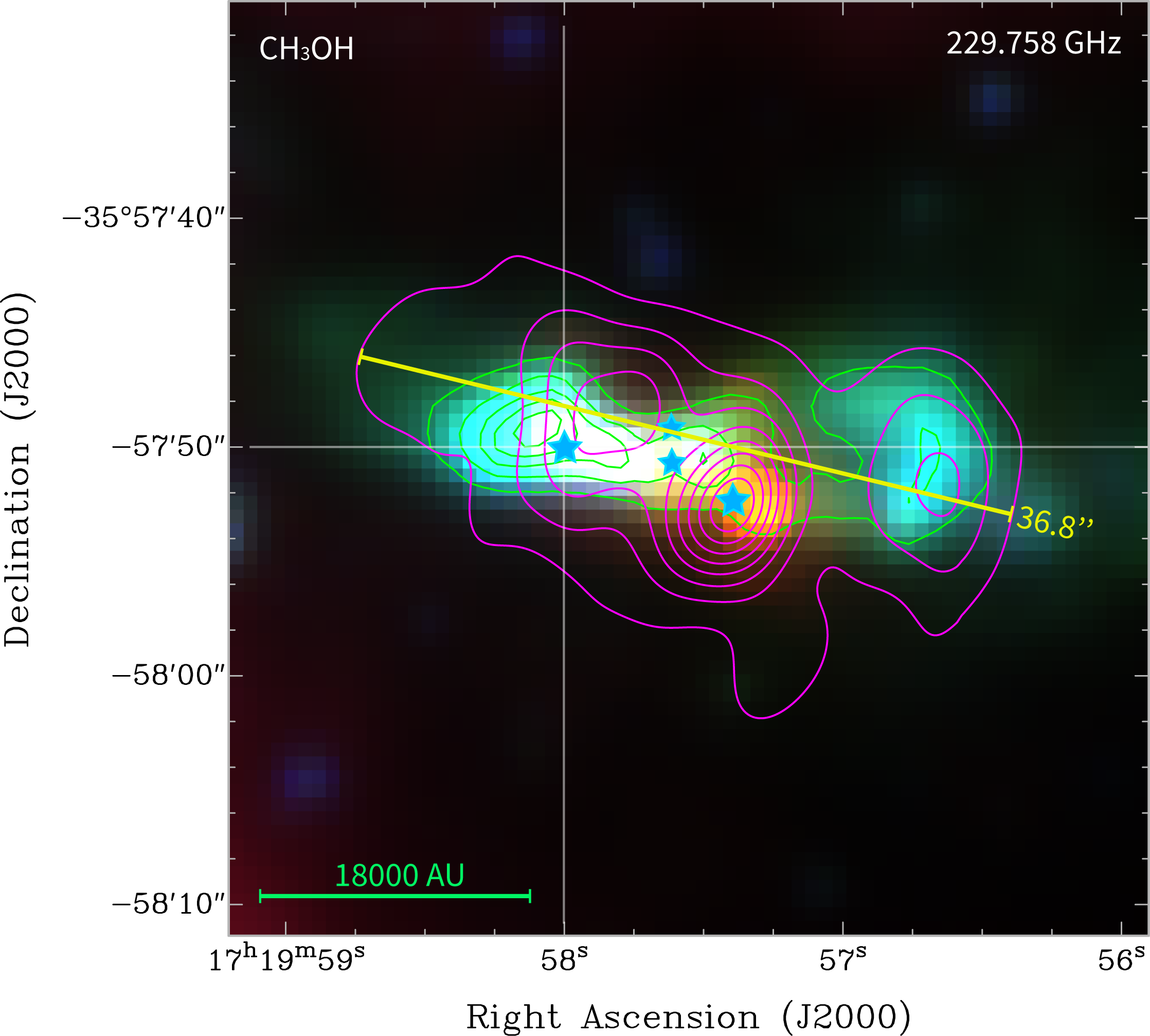} &
    \subfigimg[width=\linewidth]{c)}{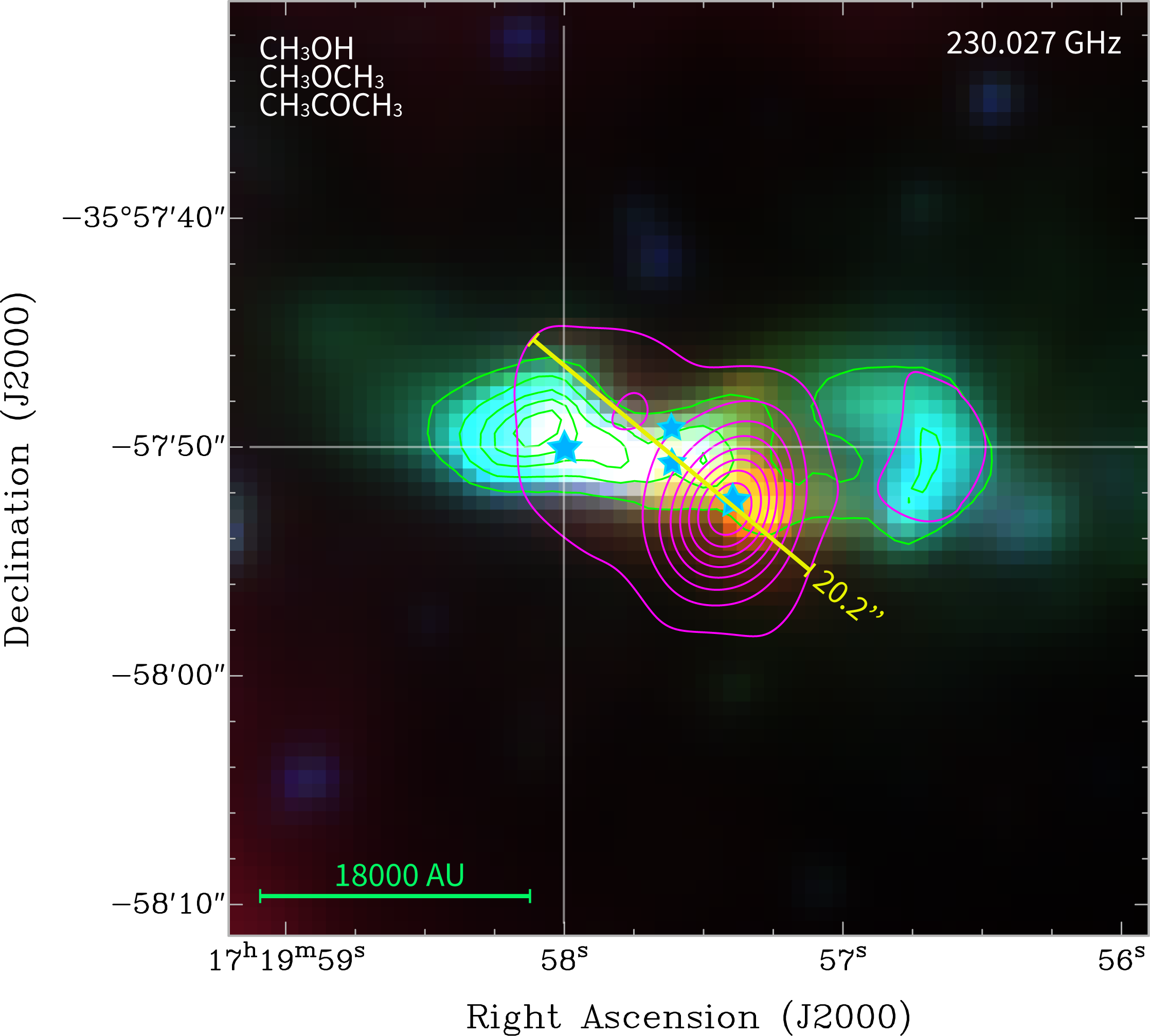} \\   
  \end{tabular}
\caption{G351 Methanol integrated emission maps at 220.078, 229.758, and 230.027 GHz. The magenta contours go from 20 to 90\%  of the emission peak with a 10\% increasing steps. These contours are displayed over a three-band color image (3.6, 4.5, and 8.0$\mu m$, coded as blue, green, and red, respectively \citep[][]{Cyganowski2008}). The star markers shows the near- to far IR confirmed cores \citep{Kraemer1999ApJ...516..817K,Hashimoto2007}}
	\label{Methanol_M0}
\end{figure*}

Our observations show that the simplest iCOM, CH$_3$OH, is located along the  $^{13}$CO HV outflow away from the KDJ4 core. This behavior is observed in two unblended transitions, at 220.080 and 229.766 GHz, and also in blended emission at 230.027 GHz (Figure \ref{Methanol_M0} a, b and c, respectively).

The CH$_3$OH emission covers an extension from 18$''$ up to 37$''$, for 220.080 and 229.766 GHz transitions, respectively. It extends along with the EGO 4.5$\mu$m emission tracing the brightest knots, e.g. the EN-A2 and the WN-A2 cores and the bow-shock structure. Nevertheless, we recall that this extended area is not symmetrical; it is enhanced preferentially in the blue-shifted emission of the $^{13}$CO HV outflow (see Figure \ref{Methanol_M0}). 

At 230.027 GHz, the methanol emission is blended with CH$_3$OCH$_3$ and CH$_3$COCH$_3$. We diagnose this mixture due to our LTE modeling, but the contribution to the emission line from these two secondary iCOMs is not as dominant as Methanol. This, together with the fact that CH$_3$OCH$_3$ and CH$_3$COCH$_3$ are expected to trace mainly the \textit{Core}, leads us to consider the extended contours of these lines are traced by methanol.  Additionally, this transition also traces the G351 \textit{Bow-shock}. This kind of methanol distribution inhabiting outflows has been reported in other objects \citep{Ospina-Zamudio2018A&A...618A.145O,Orozco_Aguilera_2018,palau2017}. 


The physical conditions at KDJ4 according to our methanol LTE modeling are displayed in Table \ref{Phy_Properties_iCOM_Table}. 
However, we point out that these values must be taken carefully, as they represent the densest and hottest region, not precisely related with the outflows, for which the density and temperature decrease. This is shown in the Methanol populated states, that for the cores can reach the highest $E_{up}$ energy levels, whereas in the outflow the populated energy levels are commonly $E_{up}<100$ K \citep{Ospina-Zamudio2018A&A...618A.145O}. 

Also, our reported physical parameters are a lower limit for the physical state of the gas, since in our interferometric observations the amount of filtered flux can vary significantly. Further single-dish observations should help to quantitatively estimate the proportion of missing flux.

\subsubsection{Vynil Cyanide (C$_2$H$_3$CN)}
\begin{figure}
	\centering  
\includegraphics[width=.80\linewidth]{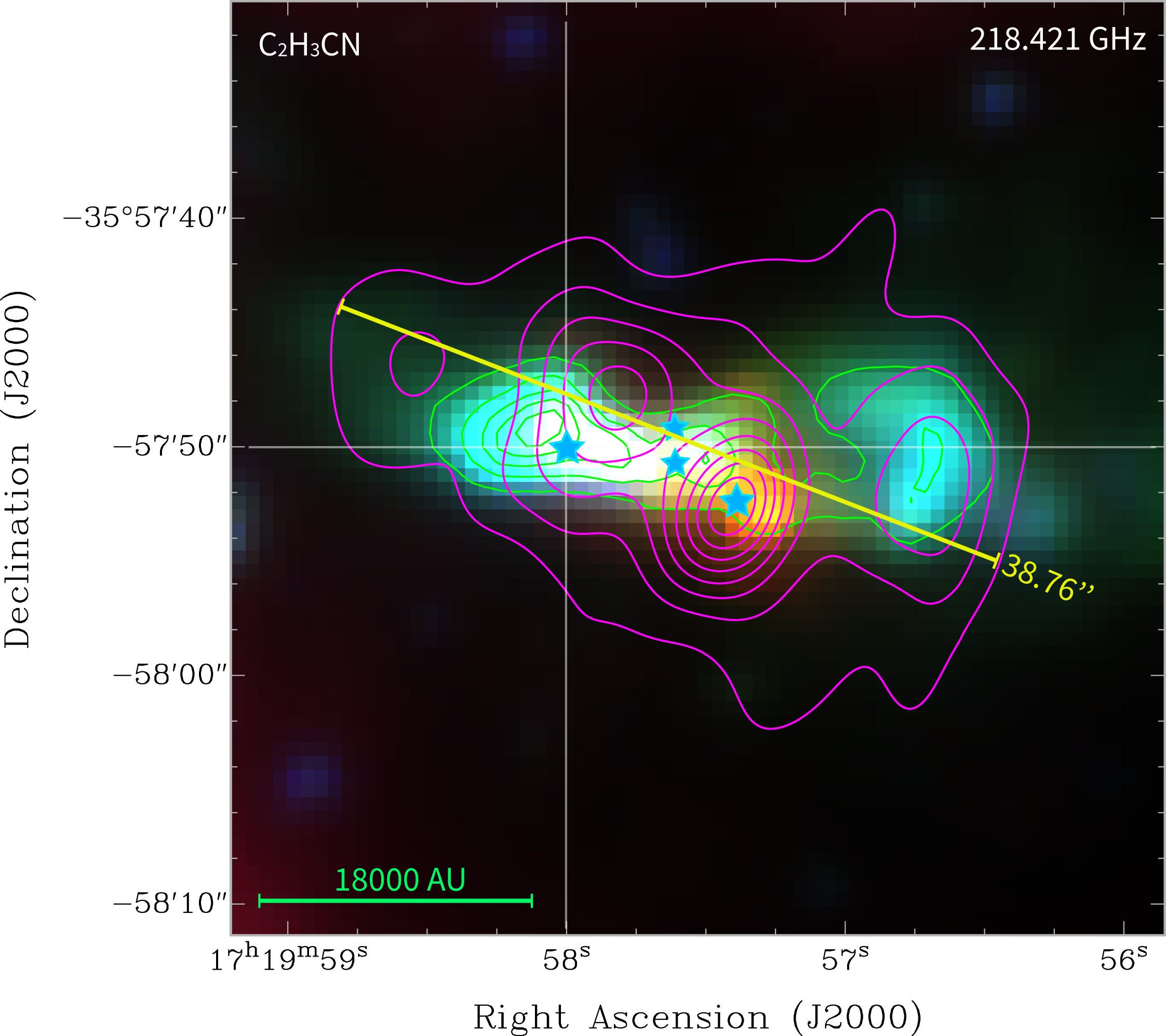}
	\caption{G351 Vinyl Cyanide integrated emission map (218.421 GHz). The contours, labels, and background image are the same as in Figure \ref{Methanol_M0}.}
	\label{Vinyl_Cyanide_M0}
\end{figure}

The most extended iCOM emission observed in G351 is Vinyl Cyanide at 218.433 GHz (38.7$''$). The emission covers all the \textit{EGO} extension, including the \textit{Bow-Shock}; however, its P.A. is more consistent with the $^{13}$CO HV outflow than with the \textit{EGO} (Figure \ref{Vinyl_Cyanide_M0}). As with the methanol emission, the vinyl cyanide traces the WN-A2 and the EN-A2 cores, but its elongation is broader than the methanol one in both axes. 

The enhancement of this molecule along with other \textit{N}-bearing species appears to be more density selective than the \textit{O}-bearing ones. This is shown as an increase of the abundance with the mass of the system by \citet{Ospina-Zamudio2018A&A...618A.145O}. Other proposed factors to produce this selectivity are the accretion rate and age of the star formation system, but there is not an established consensus yet \citep{Lykke2017A&A...597A..53L,lopez2017A&A...606A.121L, Watanabe2017ApJ...847..108W}.\\

\subsubsection{Ethyl Cyanide (C$_2$H$_5$CN) and Methyl formate (CH$_3$OCHO)} 
\begin{figure*}
  \centering
  \begin{tabular}{@{}p{0.33\linewidth}@{\quad}p{0.33\linewidth}@{}p{0.33\linewidth}}
    \subfigimg[width=\linewidth]{a)}{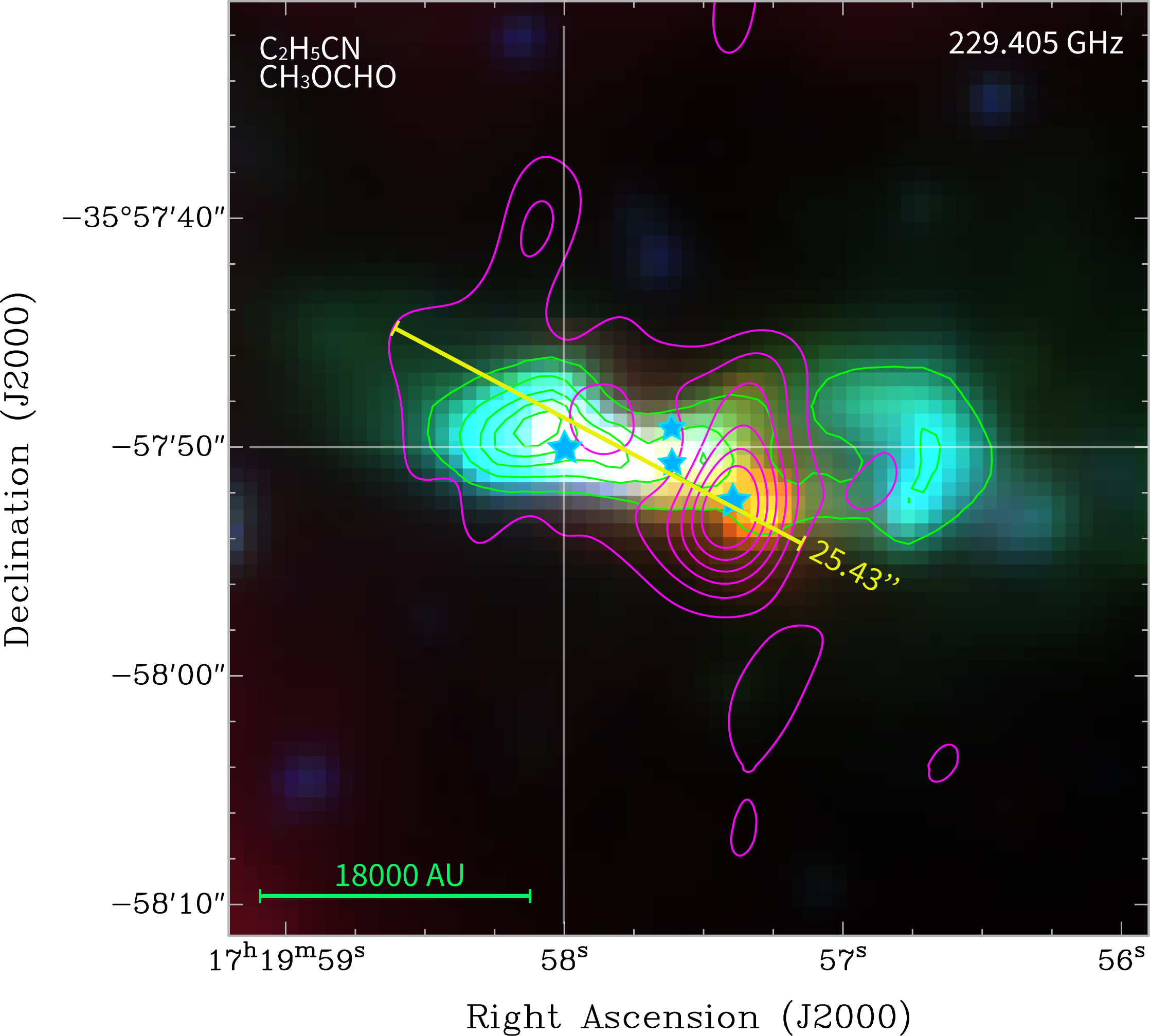} &
    \subfigimg[width=\linewidth]{b)}{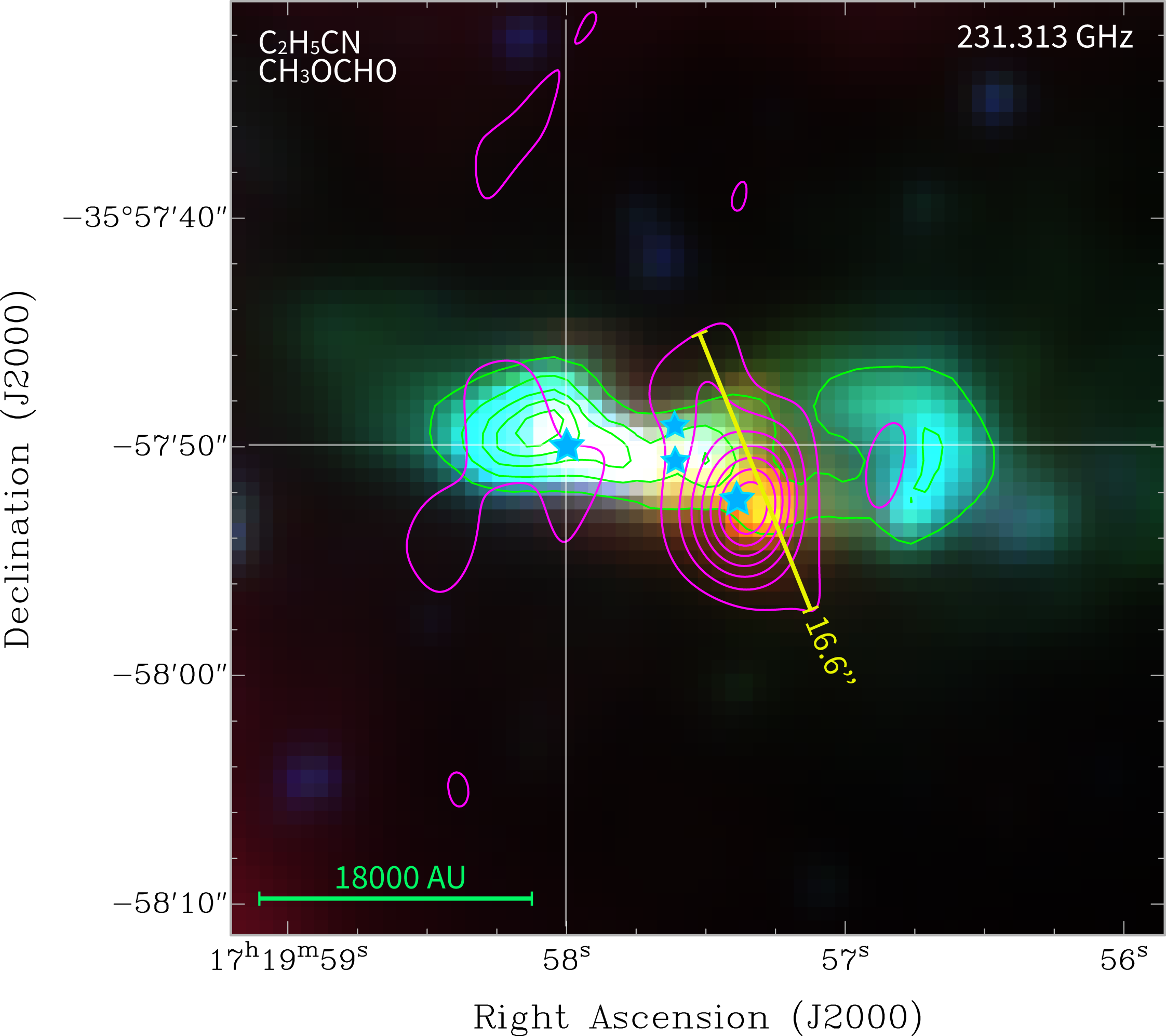} &
    \subfigimg[width=\linewidth]{c)}{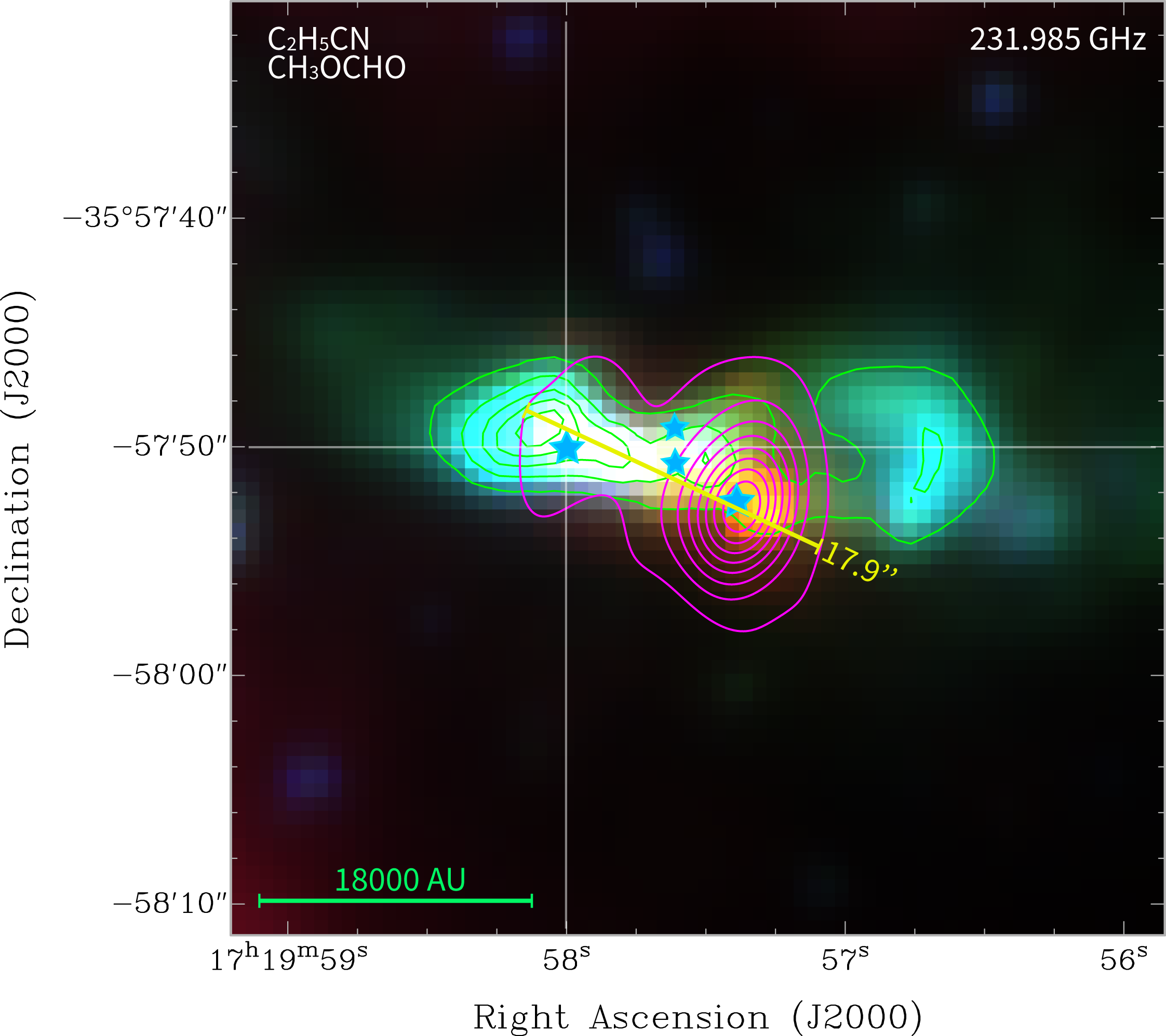} \\   
  \end{tabular}
	\caption{G351 Ethyl Cyanide and MF integrated emission maps at 229.405, 231.313, and 231.985 GHz. The contours, labels, and background image are the same as in Figure \ref{Methanol_M0}.}
	\label{Ethyl_Cyanide_M0}
\end{figure*}

In our observations, the Ethyl cyanide is observed in three emission lines 229.405, 231.313, and 231.990 GHz, but in all these cases is mixed with CH$_3$OCHO. The transition at 229.405 GHz is the more elongated one, and is tracing the blue-shifted region of the $^{13}$CO HV outflow over 25.43$''$ (Figure \ref{Ethyl_Cyanide_M0} a). 

Considering the nature of these two molecules, we speculate that the Ethyl Cyanide emission is mainly contributing to the central core emission at KDJ4, as enhancement of \textit{N}- bearing molecules has been found in massive and hot compact regions. On the other hand, Methyl Formate (MF, hereinafter) has been proven to trace also the outflowing gas knots \citep{palau2017,Orozco_Aguilera_2018}. Hence, it seems likely that the extended emission is mostly traced by MF.

The emission at 231.313 and 231.985 GHz is more compact than at 229.405 GHz and is located at the KDJ4 core. Its P.A. is more north-south and seems to agree with the HV outflow traced by SiO for over 16.6$''$. A small extension on the emission line at 231.985 GHz seems to trace the north-east direction of the $^{13}$CO HV outflow, but could be just tracing the cores (see Figure \ref{Ethyl_Cyanide_M0} c). This small lobe extends for 17.9$''$ and is tracing the EN-A2 and WN-A2 cores.


\subsubsection{Methyl formate (CH$_3$OCHO); Ethylene glycol aGg'-(CH$_2$OH)$_2$}

\begin{figure}
  \centering
  \begin{tabular}{@{}p{0.80\linewidth}}
    \subfigimg[width=\linewidth]{a)}{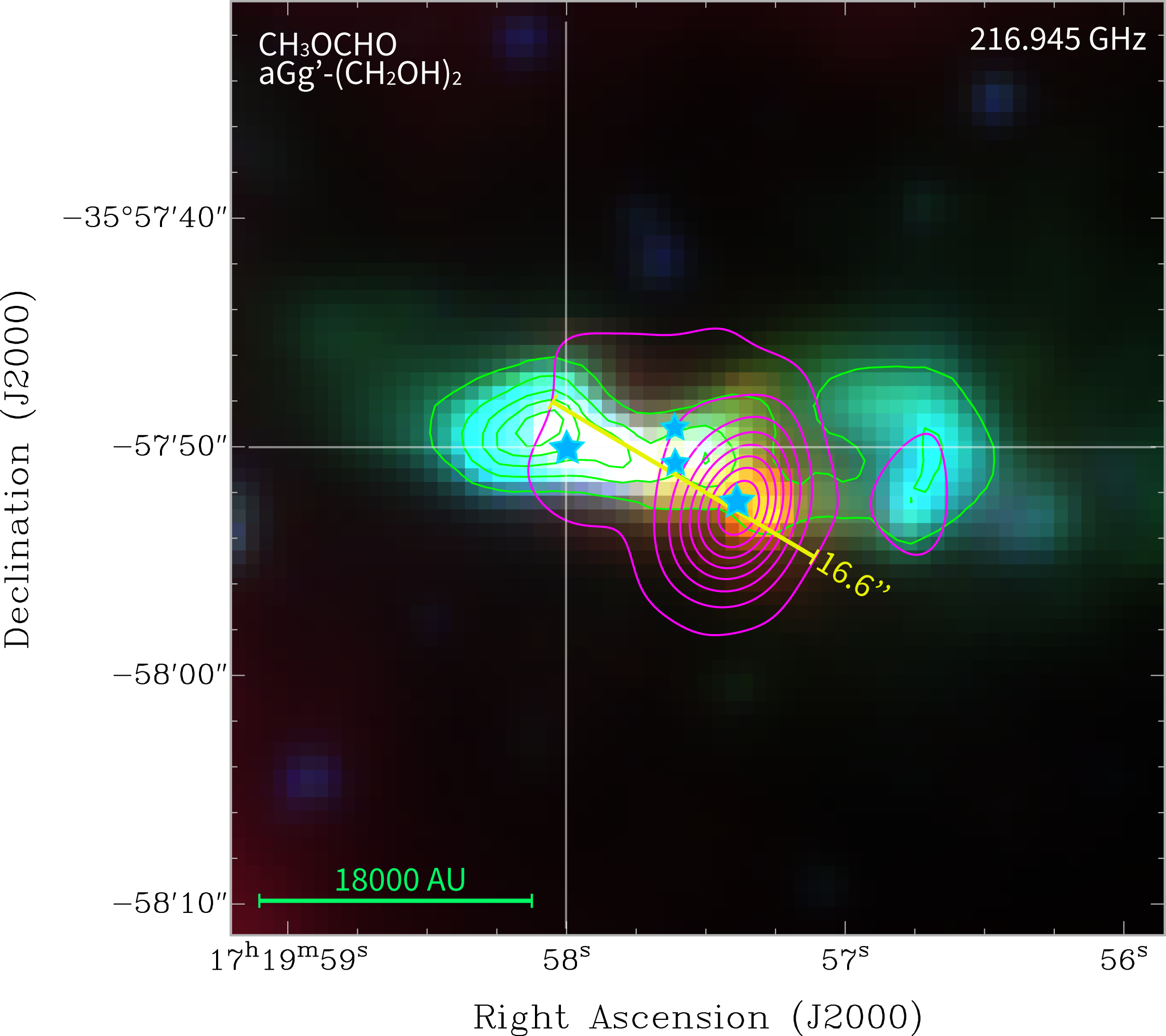} \\
    \subfigimg[width=\linewidth]{b)}{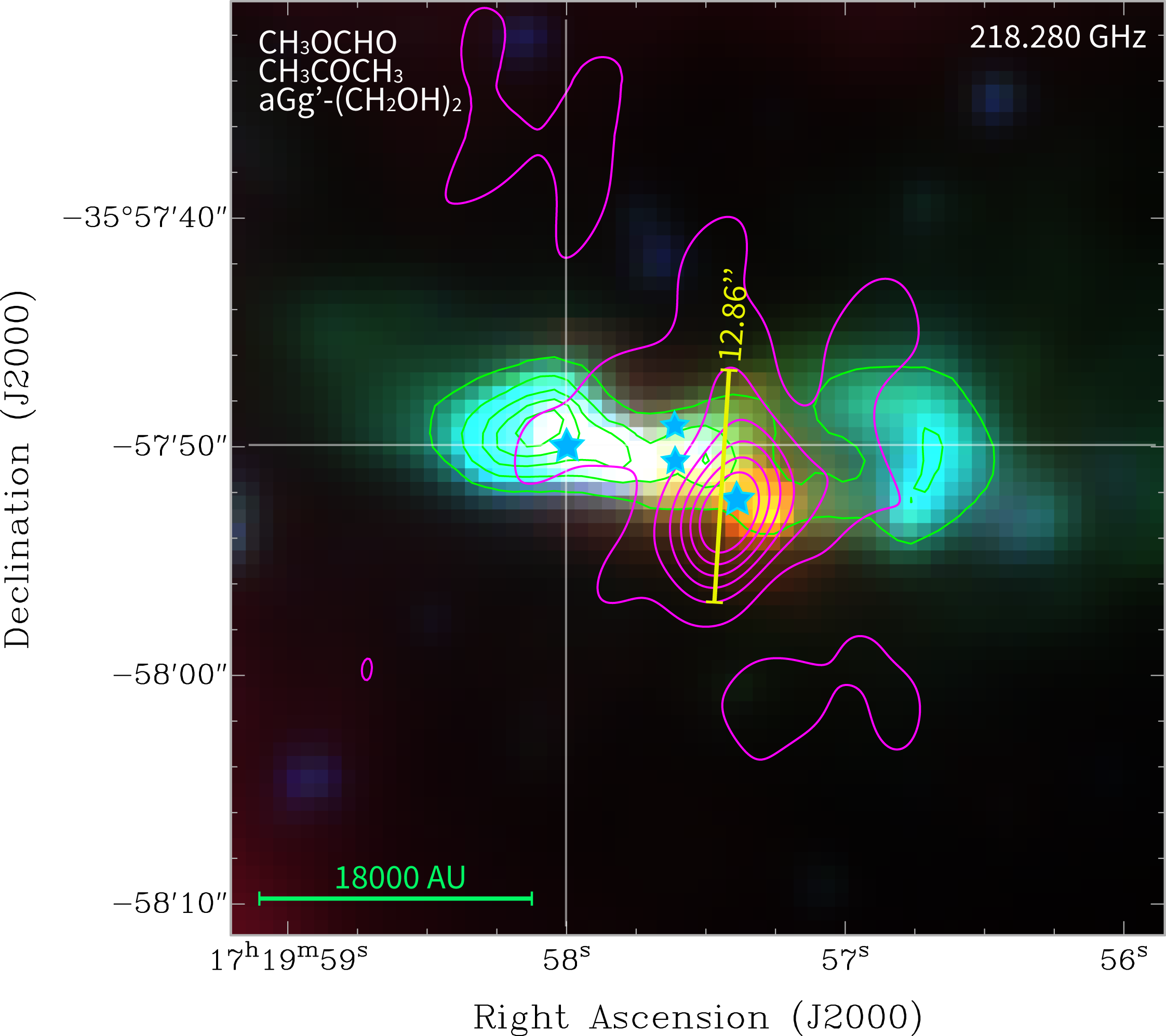} \\   
  \end{tabular}
	\caption{G351 MF and Ethylene Glycol integrated emission map at 216.945 and 218.280 GHz. The contours, labels, and background image are the same as in Figure \ref{Methanol_M0}.}
	\label{Methyl_Cyanide_M0}
\end{figure}

These two iCOMs are responsible for two elongated emissions at 216.945 and 218.280 GHz. In this case, the covered area traces similar components centered in the KDJ4 core with inner contours north-south orientated  (Figure \ref{Methyl_Cyanide_M0}). This extension is also similar to the C$_2$H$_5$CN, CH$_3$OCHO, and CH$_3$OH. Methanol and methyl formate are expected to inhabit similar regions, as methyl formate is directly produced after the processing of Methanol \citep{Ceccarelli2017ApJ...850..176C}. Also, their enhancement in outflow regions has been previously confirmed \citep{Rojas_Garcia_2022}. 



\subsubsection{Other iCOMs}
We also found several other iCOMs associated with elongated emission such as Acetone (CH$_3$COCH$_3$ at 218.281 GHz) and Dimethyl ether (CH$_3$OCH$_3$ at 230.027 GHz), but as they were blended with simpler molecules, and have high energy levels, they may be tracing the compact core. Additionally, as a result of our line identification methodology, we were able to detect several other iCOMs shuch as C$_2$H$_5$OH, NH$_2$CHO and CH$_3$CN along with the isotopologues C$_2$H$_5$CN$^{15}$, C$_2$H$_5$C$^{13}$N, CH$_3$OC$^{13}$HO, CH$_3$C$^{13}$N and C$^{13}$H$_3$CN, but as they only trace the compact core, they were excluded from this analysis. The full list of iCOMs is displayed in Table \ref{Phy_Properties_iCOM_Table}.

\begin{table}
	\centering
	\caption{Physical parameters obtained for iCOMs detected into the brightest core KDJ4. The parameters have been determined by the XCLASS LTE modeling, i.e. this quantities represent the physical parameters corresponding to one beam coverage ($\theta_{syth}=5.3'' \times 3.6''$). 
 The last column shows the molecular abundances relatives to  H$_2$ column density of \num{4.0E23}, in order to facilitate comparisons with literature values.}
	\footnotesize 
	\begin{tabular}{ccccccccccccccccc}
		\toprule
iCOM 	&T$_{rot}$	&N$_{Tot}$	&$v_{width}$	& $X$\\
specie & [K] & [cm$^{-2}$] & [kms$^{-1}$] &  \\ 
		\midrule
    aGg'-(CH$_2$OH)$_2$         	&146	&\num{1.98E+15}	&3.21	& \num{2.95E-09} \\
    C$^{13}$H$_3$CN                 & 82	&\num{2.83E+14}	&2.01	& \num{4.21E-10} \\
    C$_2$H$_3$CN                    &546	&\num{2.31E+15}	&6.04	& \num{3.44E-09} \\
    C$_2$H$_5$C$^{13}$N            	&160	&\num{1.65E+13}	&7.10	& \num{2.46E-11} \\
    C$_2$H$_5$CN$^{15}$             & 25	&\num{3.72E+16}	&4.97	& \num{5.54E-08} \\
    C$_2$H$_5$CN                    &743	&\num{8.00E+15}	&3.38	& \num{1.19E-08} \\
    C$_2$H$_5$OH                    &177	&\num{1.50E+16}	&4.00	& \num{2.23E-08} \\
    CH$_3$C$^{13}$N             	&155	&\num{2.78E+14}	&3.53	& \num{4.14E-10} \\
    CH$_3$CHO               	   &57	&\num{4.28E+14}	&1.35	& \num{6.37E-10} \\
    CH$_3$CN                     &200	&\num{4.32E+15}	&3.72	& \num{6.43E-09} \\
    CH$_3$COCH$_3$                  &286	&\num{2.20E+17}	&8.51	& \num{3.28E-07} \\
    CH$_3$OC$^{13}$HO               &221	&\num{1.55E+16}	&3.60	& \num{2.31E-08} \\
    CH$_3$OCH$_3$                   &289	&\num{3.69E+16}	&5.02	& \num{5.49E-08} \\
    CH$_3$OCHO                   &251	&\num{3.77E+16}	&4.67	& \num{5.61E-08} \\
    CH$_3$OH                     &296	&\num{1.52E+17}	&4.21	& \num{2.26E-07} \\
    NH$_2$CHO                    & 32	&\num{5.32E+14}	&4.47	& \num{7.92E-10} \\
	\bottomrule                    
	\end{tabular}\label{Phy_Properties_iCOM_Table}
\end{table}

\subsection{Velocity field maps of iCOMs tracing the bow shock}
We analyzed the moment 1 maps for the iCOMs that trace the bow-shock signature in their integrated emission maps to visually inspect the velocity gradient. It should be pointed out that due to the multiple core nature of G351, the velocity gradient field is not smooth but incoherent; therefore, the precise association of the moving gas traced by iCOMs with a particular core is not easy. 

Nevertheless, we have analyzed the velocity field mainly focused on two outflow signatures previously mentioned: the blue-shifted HV $^{13}$CO outflow, and the bow shock structure (Section \ref{subsec:Structure_g351}).

\subsubsection{Methanol}
\begin{figure*}
  \centering
  \begin{tabular}{@{}p{0.33\linewidth}@{}p{0.33\linewidth}@{}p{0.33\linewidth}}
    \subfigimg[width=\linewidth]{a)}{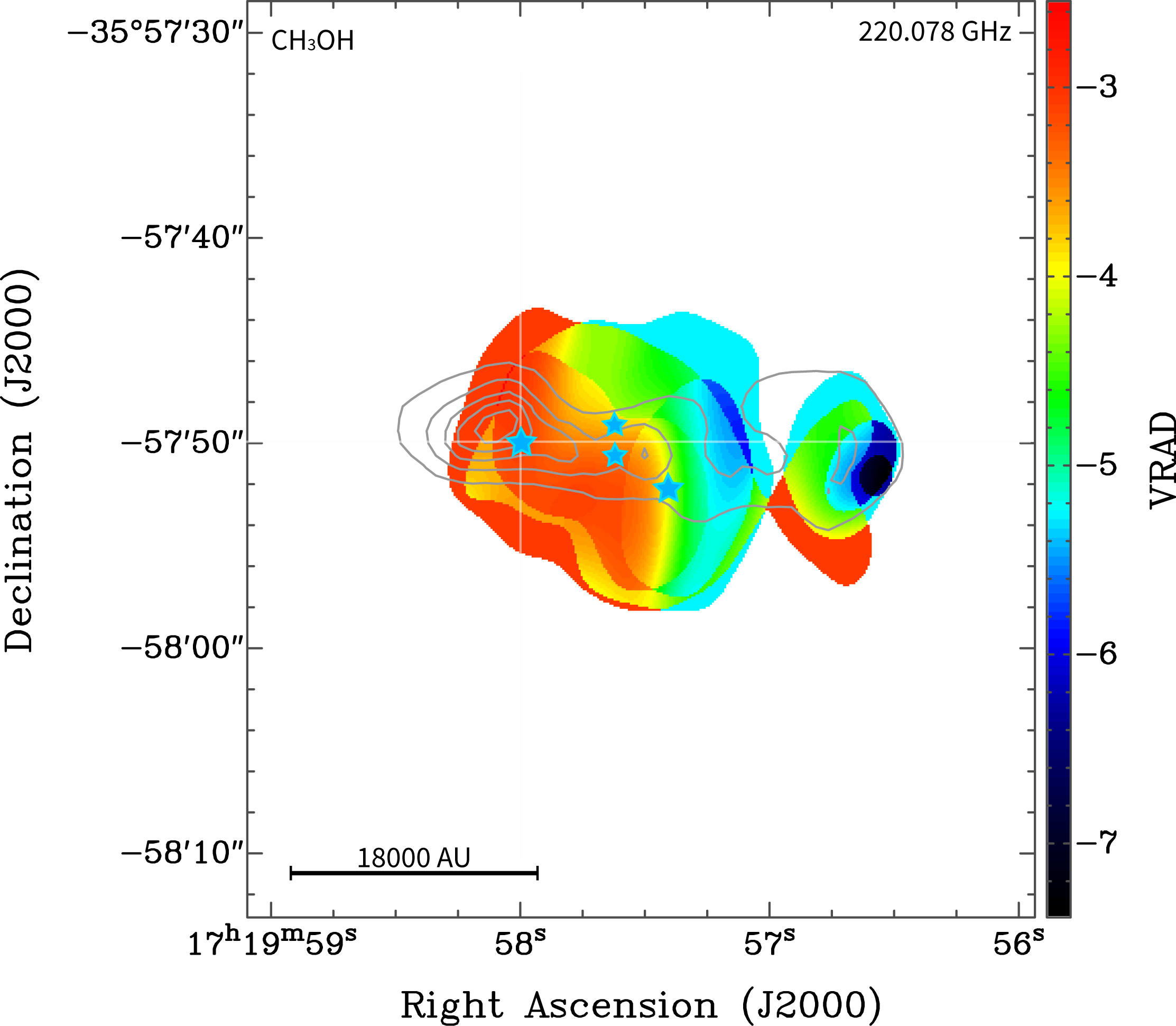} &
    \subfigimg[width=\linewidth]{b)}{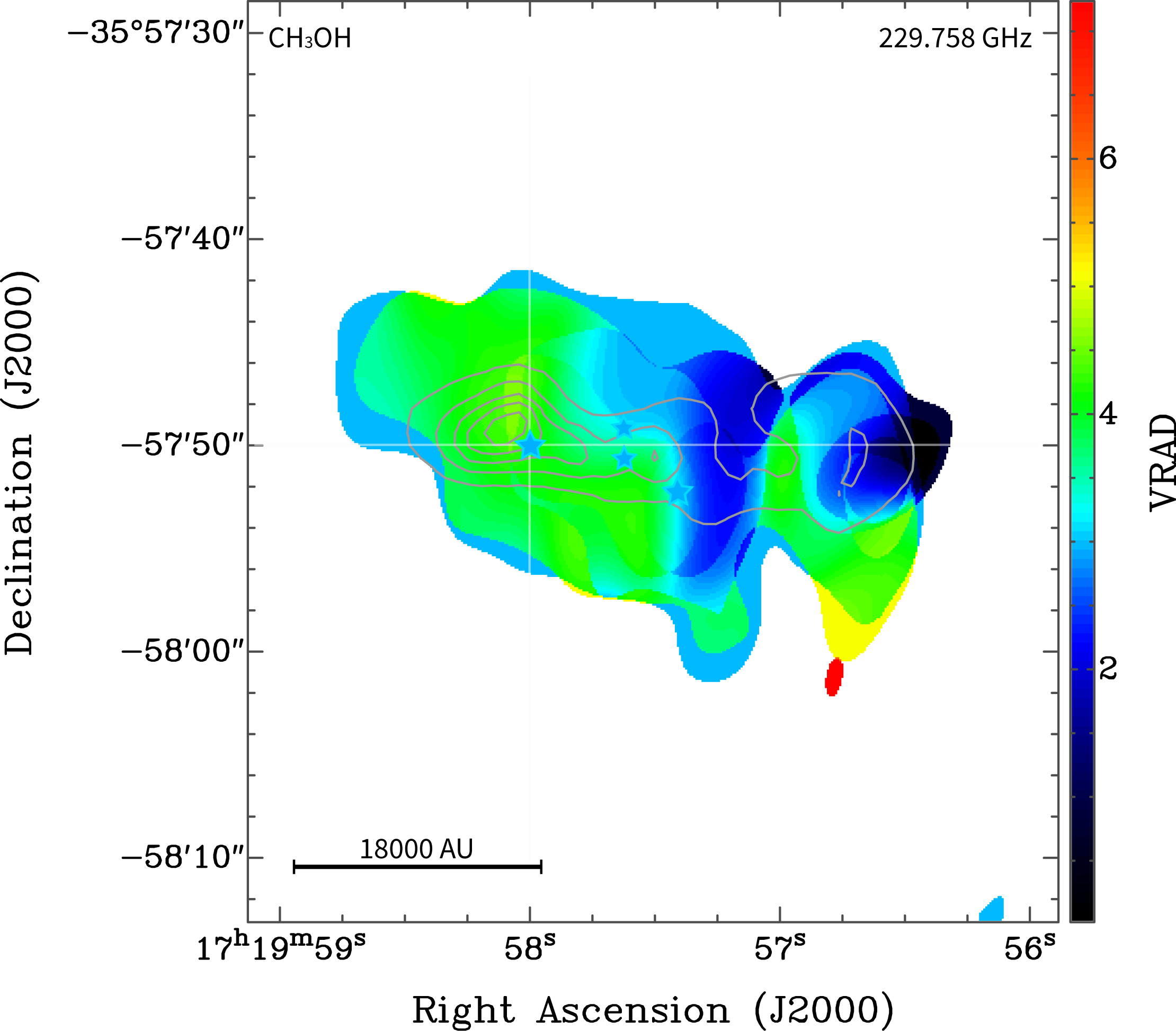} &   
    \subfigimg[width=\linewidth]{c)}{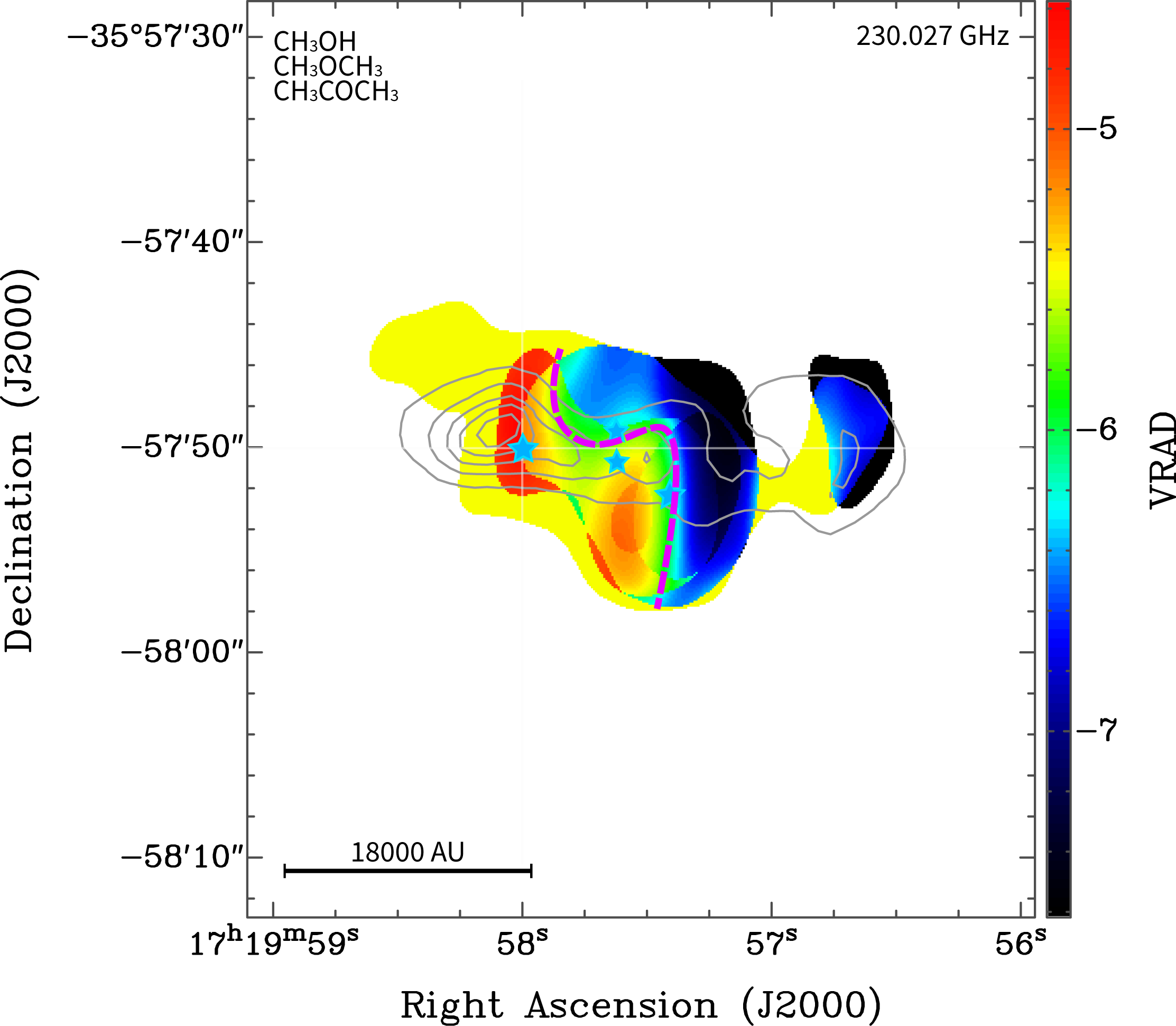} \\   
  \end{tabular}
	\caption{G351 Methanol first order moment maps at 220.078, 229.758, and 230.027 GHz. The maps were generated by Kpvslice software from KARMA using an \textit{median} algorithm with lower clips of 4$\sigma$.  The star markers show the near- to far IR confirmed cores \citep{Hashimoto2007}. In \textit{c)} the dashed magenta line highlights the rest velocity towards this blended emission line.}
	\label{Methanol_M1}
\end{figure*}

In the unblended emission lines at 220.078 and 229.758 GHz, methanol M1 maps show the blue-shifted emission located just over the bow shock structure and pointing to the west, with a center near to WN-A2; whereas the red-shifted emission spreads from east to west (Figure \ref{Methanol_M1} a and b). The velocity field traced by Methanol at 230.027 GHz 
shows an \textit{"S"} shape (dashed magenta line in Figure \ref{Methanol_M1} c), which agrees with the location of the near- to far-IR cores WN-A2 and KDJ4. This phenomenon may be caused by the secondary emitting iCOMs blended in this emission (CH$_3$OCH$_3$ and  CH$_3$COCH$_3$), which are more related to compact emission. Therefore, the velocity field could be tracing the location of the innermost regions of the compact cores, whereas the other maps only traced by Methanol show a wider emission.

Around the bow shock structure, we observe a gradient from red- to blue-shifted emission, going from south to north, with a velocity width of $|V_{Max}-V_{LSR}| < 4$ kms$^{-1}$ for the lowest frequency methanol line (Fig. 10a). Due to its location and the non-correspondence with the mm continuum peak, we speculate that the observed gradient could be related to the bow shock itself and the nearby SiO outflow. In the case of the 229 GHz line (Fig. 10b), the south to north gradient is more clearly related to the SiO outflow (see, in comparison, Figure \ref{G351_16_CO-SiO}). 

Another signature in the Methanol emission is that its velocity gradient is not coherent along the path of the HV outflow. This could indicate that methanol is tracing the low-velocity component of the HV outflow. During this exchange of kinetic energy, the reactions are boosted only in the densest layers and chemically richest knots, but not all the way along the outflow path (as the simplest molecules). In these places, the increase of temperature and density trigger extra iCOM emission. This  phenomenon also explains the low velocity in the iCOMs at these locations.


\subsubsection{Methyl formate}
Another iCOM which seems to trace the bow shock structure is MF at 216.945 GHz. Nevertheless, this emission is contaminated with aGg-(CH$_2$OH)$_2$, but according to our LTE modeling the main contributor for this emission should be MF. 

\begin{figure}
	\centering  
\includegraphics[width=0.80\linewidth]{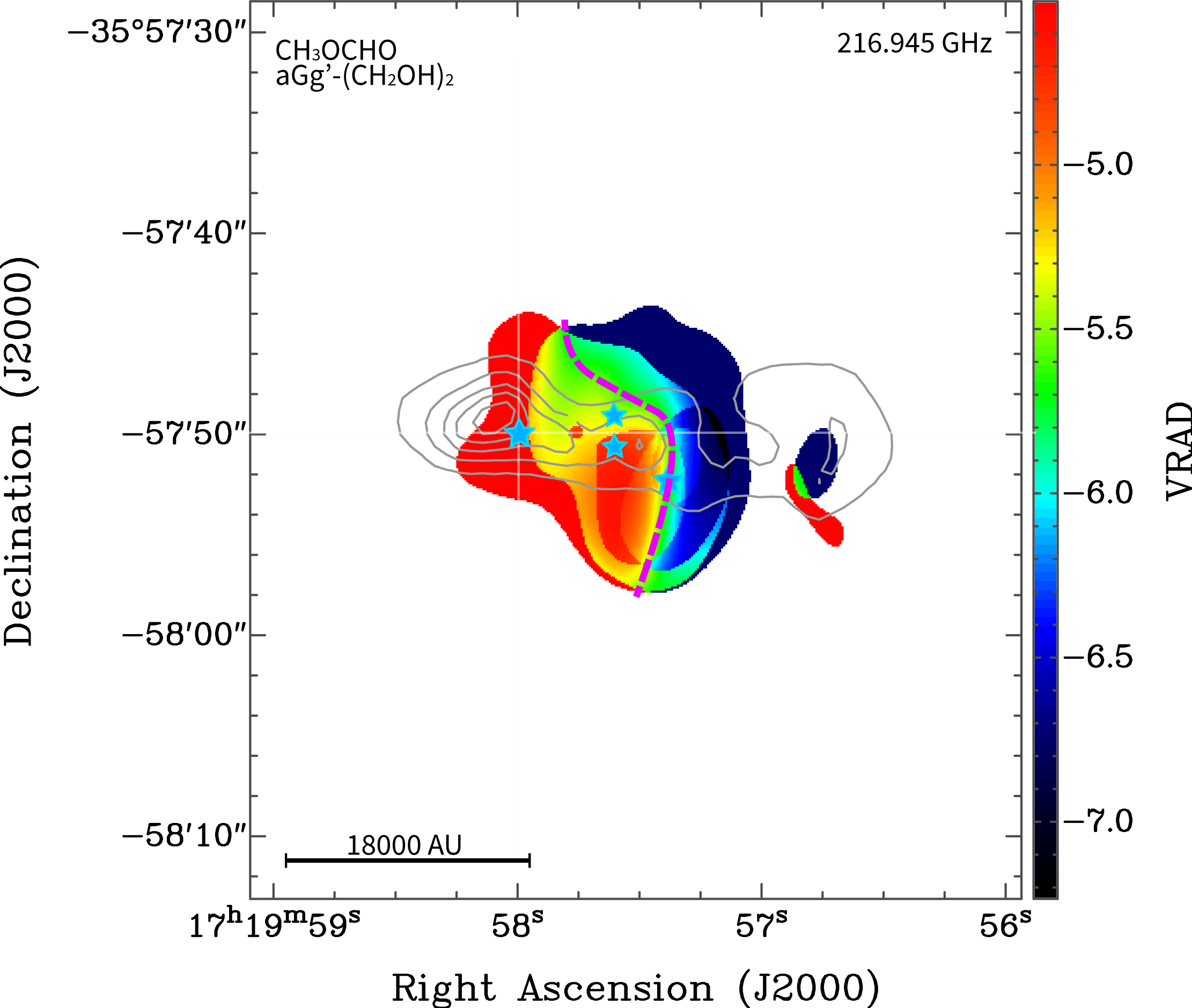}
	\caption{G351 MF first order moment map at 216.945 GHz. The map generation and the mark labels are the same as in Figure \ref{Methanol_M1}}
	\label{MF_M1 maps}
\end{figure}

As in the M1 map of contaminated Methanol, this blended emission shows an \textit{"S"} shape in the closest vicinity to the near- to far-IR cores (Figure \ref{MF_M1 maps}). Also, the bow shock gradient shows a red- to blue-shifted emission, going from south to north, but as we previously mentioned for the case of methanol, this could be due to the bow shock itself and the nearby SiO outflow.

\subsubsection{Vinyl Cyanide}
\begin{figure}
	\centering  
\includegraphics[width=0.80\linewidth]{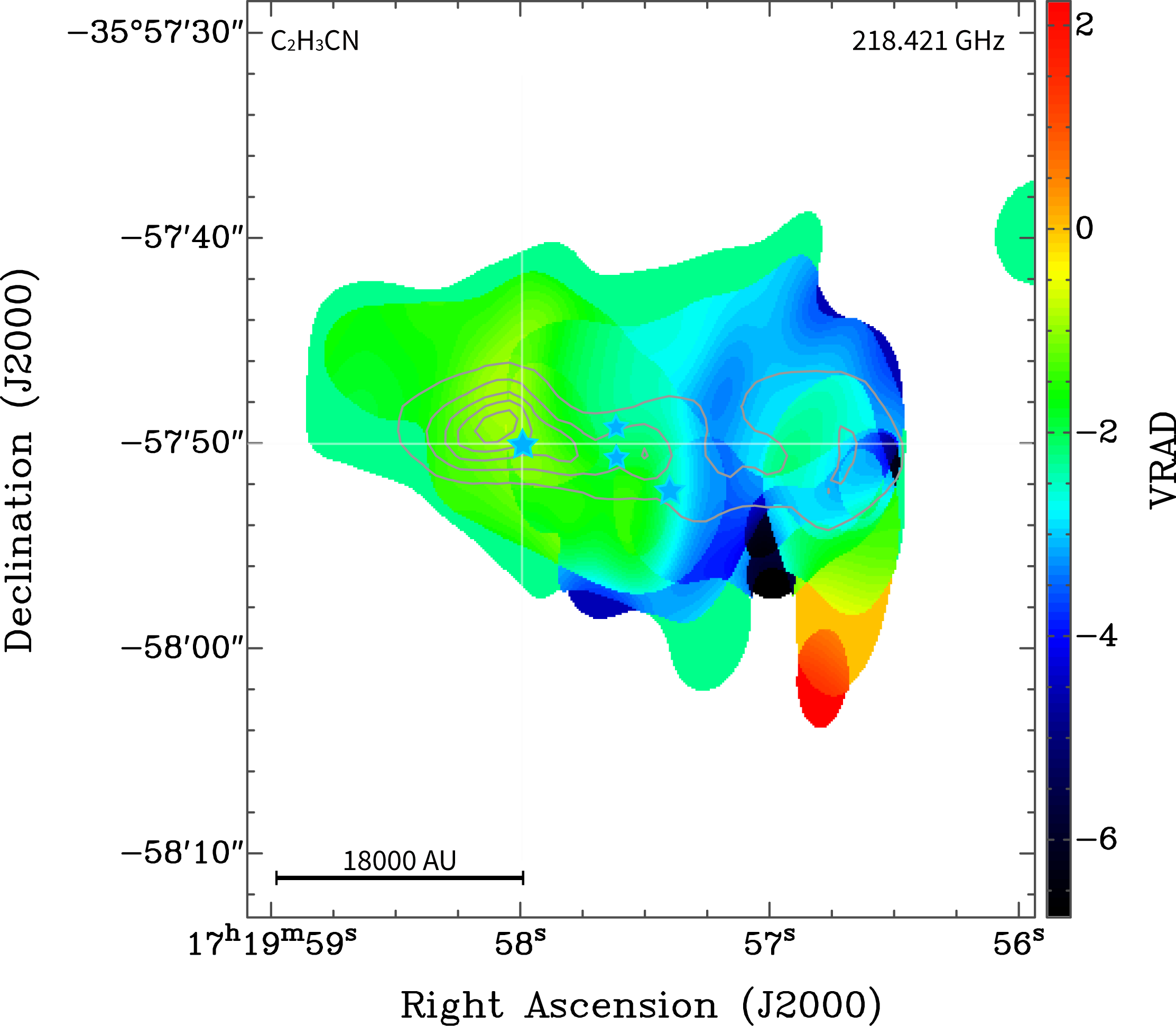}
	\caption{G351 VC first order moment map at 218.421 GHz. The map generation and the mark labels are the same as in Figure \ref{Methanol_M1}}
	\label{VC_M1_maps}
\end{figure}

The vinyl cyanide velocity field is mostly close to rest velocity, in comparison to other iCOMs. This emission is extended east to west, covering and spreading further away from the extension of the EGO, and also tracing the extension of the HV $^{13}$CO outflow (Figure \ref{VC_M1_maps}). The maximum red- and blue-shifted velocities show agreement with the south-west peak emission of the red- and blue-shifted emission of the SiO HV outflow. As this emission is elongated to the south-west from the KDJ4 core, and has a wide extension, the velocity maps overlap with the bow-shock, preventing its association with one structure or another.

\subsection{Molecular abundances}

To put in context our results along the star formation scenario from low to high mass, we compute the relative abundances of iCOMs with respect to H$_2$, using the computed column densities. The H$_2$ column density was computed using the \citet{Hernandez2014ApJ...786...38H} procedure, given the similar frequency ranges. The first assumption is optically thin dust emission and a constant gas-to-dust mass ratio (100, in this case). The opacity per unit of dust ($\kappa_v$) goes from 0.2 to 3.0 at 1.3 mm, and it is scaled with $\nu^{\beta}$. Following  \citet{Hernandez2014ApJ...786...38H}, we took $\beta=1.5$, leading to $\kappa_{1.3mm}=0.74$ cm$^{-2}$g$^{-1}$. Therefore, the column density of molecular hydrogen could be written as

\begin{equation}
    \Bigg[\frac{N_{H_{2}}}{cm^{-2}}\Bigg]=\frac{\num{2.35E16}}{\theta^2}\Bigg[\frac{S_{\nu}}{Jy}\Bigg]\Bigg[\frac{T}{K}\Bigg]^{-1}
\end{equation}\vspace{1pt}

Here $S_{v}$ is the flux density, $\theta$ is the source size in rad, and $T$ is the dust temperature. In this case, we take the rotational temperatures average value from the optically thin isotopolgues of $^{13}$CO ($T_{Rot}=24.88$ K) and CO$^{18}$ ($T_{Rot}=24.35$ K), resulting in a $T_{Rot}\sim25$ K. This temperature is similar to the one assumed by \citet{Juarez17} for the north and southeast cores (that is also compatible with to the average temperature in protostellar cores $T_{Rot}\sim30$ K). The continuum was taken from our continuum data, 464.16 mJy beam$^{-1}$. After these considerations, we obtain a $N_{H_{2}} = \num{6.72E23}$ cm$^{-2}$.
The column density ratios are listed in Table \ref{Abundancies_iCOM_Table}.



\begin{landscape}
\begin{table}
	\centering
	\caption{Molecular abundances with respect to H$_2$ measured from low to high mass SFRs reported in literature along our own results for G351 into the brightest core KDJ4. The sources are ordered by increasing luminosity from left to right.  ($^a$) López-Sepulcre et al.
(2017), Taquet et al. (2015), ($^b$) Jørgensen et al. (2016), Coutens et al. (2016), Lykke et al. (2017), Calcutt et al. (2018), ($^c$) \citet{Ospina-Zamudio2018A&A...618A.145O}, ($^d$) Fuente et al. (2014),
($^e$) Beuther et al. (2009), ($^f$) Pagani et al. (2017), Beuther et al. (2009), ($^g$) \citet{palau2017}, ($^h$) Belloche et al. (2013). (*) Obtained from the $^{13}$C isotopologue. (+) Average value from the reported range.} 
	\begin{tabular}{C{1.7cm}C{1.4cm}C{1.4cm}C{1.4cm}C{1.4cm}C{1.4cm}C{1.4cm}C{1.4cm}C{1.4cm}C{1.4cm}C{1.4cm}C{1.4cm}}
		\toprule
\multirow{6}{1.3cm}{iCOM specie}	& \multicolumn{9}{c}{$X$} \\ \cmidrule(r){2-12}
 &\multicolumn{3}{c}{Low Mass}   &\multicolumn{2}{c}{Intermediate Mass} &\multicolumn{5}{c}{High Mass} \\
 \cmidrule(r){2-4}\cmidrule(r){5-6}\cmidrule(r){7-12}
               & IRAS- 4A2$^a$       & IRAS- 16293B$^b$ & IRAS- 2A$^a$  & Cep E-A$^c$        & NGC- 7129$^d$ & G29.96$^e$  & Orion- Kl$^f$     & IRAS 20126+4104$^g$ \textit{disc} & IRAS 20126+4104$^g$  \textit{outf}  & \textbf{G351.16 +0.70} & SgrB2 (N1)$^h$ \\
               		\midrule                                                                                                                                              
CH$_3$OH       & \num{4.3E-7}        & \num{1.7E-6}     & \num{1.0E-6}  & \num{1.2E-6}*      & \num{1.0e-6}*  & \num{6.7E-8}  & \num{9.3E-8}   &    \num{9.3e-7}     &    \num{1.4e-6}     & \num{2.3E-07}    & \num{1.4E-6}    \\
CH$_3$OCHO     & \num{1.1E-8}        & \num{3.3E-8}     & \num{1.6E-8}  & \num{2.6E-8}       & \num{2.0e-8}   & \num{1.3E-8}  & \num{4.4E-9}   &    \num{9.3e-8}     &    \num{1.0e-7}     & \num{5.6E-08}      & \num{3.4E-8}    \\
CH$_3$CHO      & \num{4.3E-9}$^+$    & \num{5.8E-9}     &      -------- & \num{1.1E-8}       & \num{2.0e-9}   & --------      & \num{1.3E-11}  &    --------         &    --------         & \num{6.4E-10}     & \num{1.1E-8}    \\
CH$_3$OCH$_3$  & \num{1.0E-8}        & --------         & \num{1.0E-8}  & \num{3.6E-8}       & \num{1.0e-8}   & \num{3.3E-8}  & \num{2.6E-9}   &    \num{3.9e-8}     &    \num{1.5e-7}     & \num{5.5E-08}      & \num{1.5E-7}    \\
CH$_3$COCH$_3$ & --------            & \num{1.4E-9}     & --------      & \num{5.4E-9}$^+$   & \num{5.0e-10}  & --------      &  --------      &    \num{9.7e-8}     &    \num{3.2e-8}     & \num{3.3E-07}    & \num{1.2E-8}    \\
NH$_2$CHO      & \num{4.0E-10}$^+$   & \num{8.3E-10}    & \num{2.4E-9}  & \num{4.5E-10}      & --------       & --------      & \num{1.9E-10}  &    ---------        &    ---------        & \num{7.9E-10}     & \num{1.0E-7}    \\
CH$_3$CN       & \num{8.5E-10}$^+$   & \num{3.3E-9}     & \num{4.0E-9}  & \num{3.5E-9}       & \num{6.0E-9}*  & \num{1.7E-9}  & \num{8.5E-10}  &    \num{7.7e-9}     &    \num{7.2e-9}     & \num{6.4E-09}     & \num{1.5E-7}    \\
C$_2$H$_5$CN   & \num{3.7E-10}$^+$   & \num{3.0E-10}    & \num{3.0E-10} & \num{7.8E-10}$^+$  & \num{1.4E-9}   & \num{1.7E-9}  & \num{1.3E-9}   &    --------         &    --------         & \num{1.2E-08}   & \num{1.4E-7}    \\
	\bottomrule                    
	\end{tabular}\label{Abundancies_iCOM_Table}
\end{table}
\end{landscape}

As \citet{Ospina-Zamudio2018A&A...618A.145O} have previously shown, \textit{N}- bearing iCOMs seem to have a clear increasing abundance with the mass of the system, with a three order of magnitude increase from low to high mass regimes, observed in CH$_3$CN and C$_2$H$_5$CN species. We took the compiled information of \cite{Ospina-Zamudio2018A&A...618A.145O} and complemented it with the iCOM abundances of \cite{palau2017} and in this context, our object \textit{N}-bearing abundances fits with previous results.

Additionally, we also found a two order of magnitude enhancement of the \textit{O}-bearing CH$_3$COCH$_3$ molecule from low to high mass regimes. However, a comparison of this iCOM is hard to make, as it is only reported for one low-mass, two intermediate mass and three high-mass SFR in our reviewed sample. 

\section{Summary and conclusions}\label{SEC:summaryandconclusions}
G351 shows a peculiar outflow morphology based on the CO, SiO, and mid-IR images. The high-velocity CO emission traces a north-south outflow, most prominent in the interferometric CO map, and without a mid-IR counterpart. Additionally, a jet-like structure that dominates in the mid-IR at 8.0 $\mu$m, is mostly traced by the SiO emission and the high-velocity CO emission.

The intermediate- and high-velocity CO emission overlaps with the main mid-IR structures. In particular, the intermediate-velocity blue-shifted emission traces the west lobe that is prominent in {\it green excess}. The red-shifted intermediate-velocity emission has only a mid-IR counterpart towards its inner region. On the other hand, as revealed more clearly in the CO emission, the high-velocity blue- and red-shifted emission does not show a mid-IR counter part, and indeed such high-velocity emission seems to trace an outflow perpendicular to the mid-IR east-west outflow. Interestingly, the SiO emission that is only detected at low velocities, has been found clearly tracing the jet-like structure revealed by the 8$\mu$m emission. In addition, the SiO emission may be tracing a third outflow structure to the south of the main western lobe, which also lacks a counterpart in the mid-IR. 

Our data shows an iCOM rich environment toward G351. This enrichment is, as expected, mainly tracing the brightest core, previously associated with the dust continuum peak KDJ4 \citep{Hashimoto2007}. Away from it, most of the iCOM emission decreases drastically. Nevertheless, some of these molecules have proven to trace thinner gas which we have associated with the outflows, as their coverage area shows association with the observed bipolar structures. This was also verified in the first moment maps. 

The iCOMs tracing the outflow are mostly \textit{O}-bearing, and are also mainly tracing the HV ($\sim$ 20 kms$^{-1}$) blue-shifted lobe. This lobe is also revealed by 4.5 and 8.0$\mu$m emission in the near-IR IRAC bands, and has been previously related to two dust continuum peaks \citep{Hashimoto2007}. This fact also indicates that this iCOM extended emission could be tracing local emission from the nearby cores in their slightly shifted $V_{LSR}$. Nevertheless, some iCOMs trace the 4.5$\mu$m bow shock structure, a region which is not related to any dust continuum and is separated from KDJ4 by more than 7$''$ (9000 AU) and therefore disconnected from the core emission.  

Due to the wide angular bipolar outflow behavior, G351 could be an example of a \textit{Circulation Flow} \citep{Arce07}, in which the green fuzzy emission is more related to the Low Velocity and High-Velocity gas $|V_{Max}-V_{LSR}| < 20$ kms$^{-1}$, whereas the Extremely High velocity ($|V_{Max}-V_{LSR}| \geq 20$ kms$^{-1}$) emission have a near-IR counterpart just in their southern blue-shifted and in their norther red shifted- lobe. Another plausible option is that there is a precessing jet.

Not only do the \textit{N}-bearing species have a large discrepancy along the Intermediate-Mass vs High mass  Star Formation scenario, but also we found two orders of magnitude difference in column density in the \textit{O-}bearing species Acetone (CH$_3$COCH$_3$) from low to high mass regime objects. Also, our detection is one order of magnitude higher in T$_{Rot}$, pointing to the sensitivity of this molecule to changes in temperature, which are expected in the inner regions of these more massive cores.


All things considered, this long-scale outflow could give more hints in the description of this wide-angle outflow and its several lobes and the role of multiple cores in HMSFRs. 

\section*{Acknowledgements}

 We thank the support of researchers from the National Institute of Astrophysics, Optics and Electronics (INAOE), the Large Millimeter Telescope (LMT) and the Institute of Radio Astronomy and Astrophysics (IRyA) during the development of this investigation. We are also grateful to the Mexican funding agency Consejo Nacional de Ciencia y Tecnología (CONACYT) for its support to the INAOE's graduate program in Astrophysics. A.P. acknowledges financial support from the UNAM-PAPIIT IN111421 and IG100223 grants, the Sistema Nacional de Investigadores of CONACyT, and from the CONACyT project number 86372 of the `Ciencia de Frontera 2019’ program, entitled `Citlalc\'oatl: A multiscale study at the new frontier of the formation and early evolution of stars and planetary systems’, M\'exico. 

\section*{Data Availability}
The data underlying this article are available in  the article and the datasets were derived from sources in the public domain: \href{https://irsa.ipac.caltech.edu/data/SPITZER/GLIMPSE/index_cutouts.html}{https://irsa.ipac.caltech.edu/data/SPITZER/-GLIMPSE/index\_cutouts.html}, \href{http://atlasgal.mpifr-bonn.mpg.de/cgi-bin/ATLASGAL\_DATABASE.cgi}{http://atlasgal.mpifr-bonn.mpg.de/cgi-bin/ATLASGAL\_DATABASE.cgi} and \href{https://lweb.cfa.harvard.edu/cgi-bin/sma/smaarch.pl}{https://lweb.cfa.harvard.edu/cgi-bin/sma/smaarch.pl}.
 



\bibliographystyle{mnras}
\bibliography{MAIN} 



\appendix

\section{Full spectra with labeled iCOMs emission toward G351}

\begin{figure*}
	\centering  
	\includegraphics[width=\linewidth]{MHz_Tmb_216_218.dat_2D-01.png}
	\includegraphics[width=\linewidth]{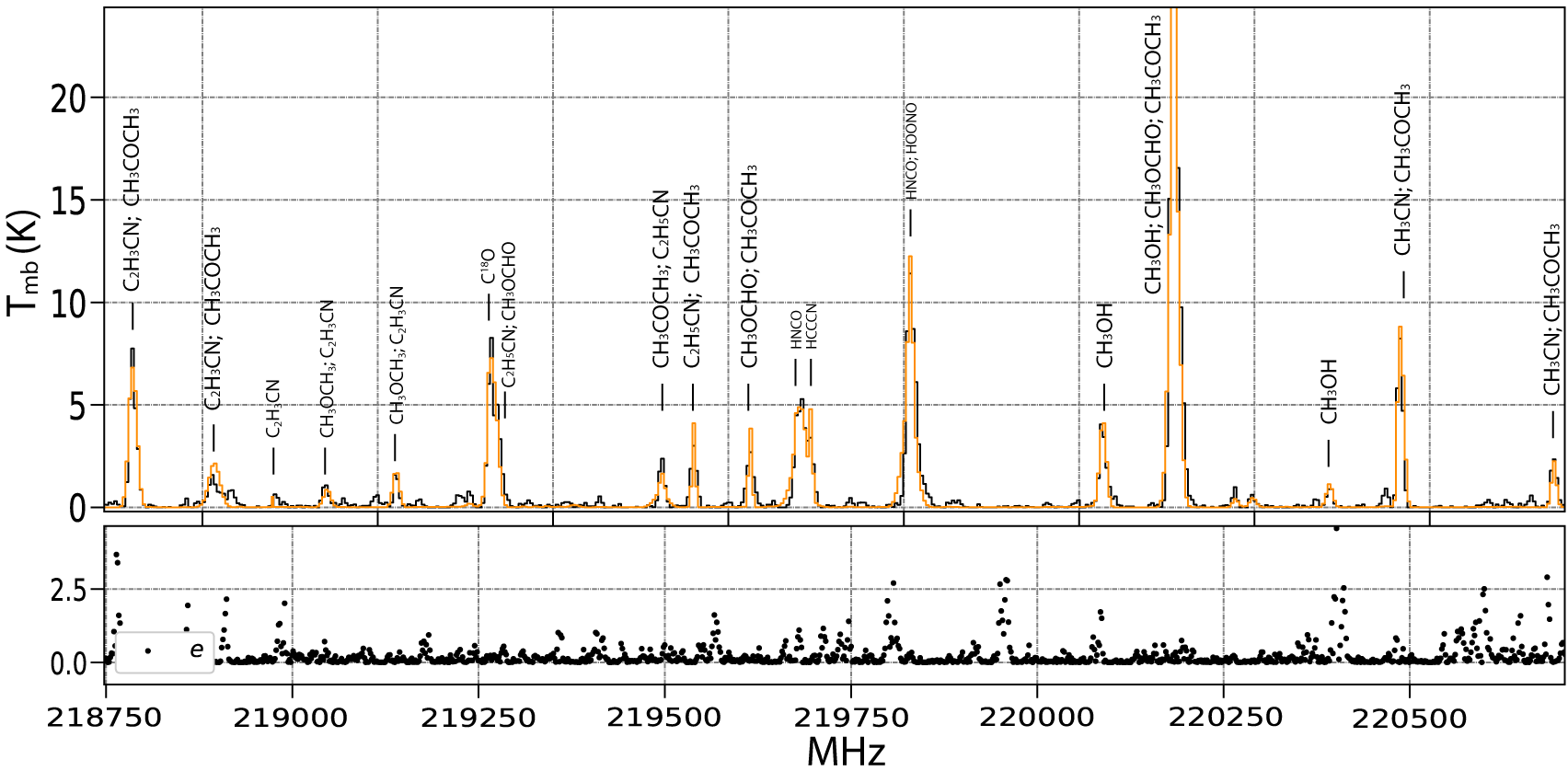}
	\caption{G351 spectra with identified iCOMs (labeled) towards their brightest core KDJ4 and their corresponding fit modeling using XCLASS software. The \textit{Black} solid line shows the observed spectra, \textit{Orange} solid line shows the modeled spectra obtained using XCLASS software. In the lower subplot, we show the residuals ($e = I_{model}-I_{obs}$).}
	\label{216-all}
\end{figure*}
\begin{figure*}
	\ContinuedFloat
\centering 
	\includegraphics[width=\linewidth]{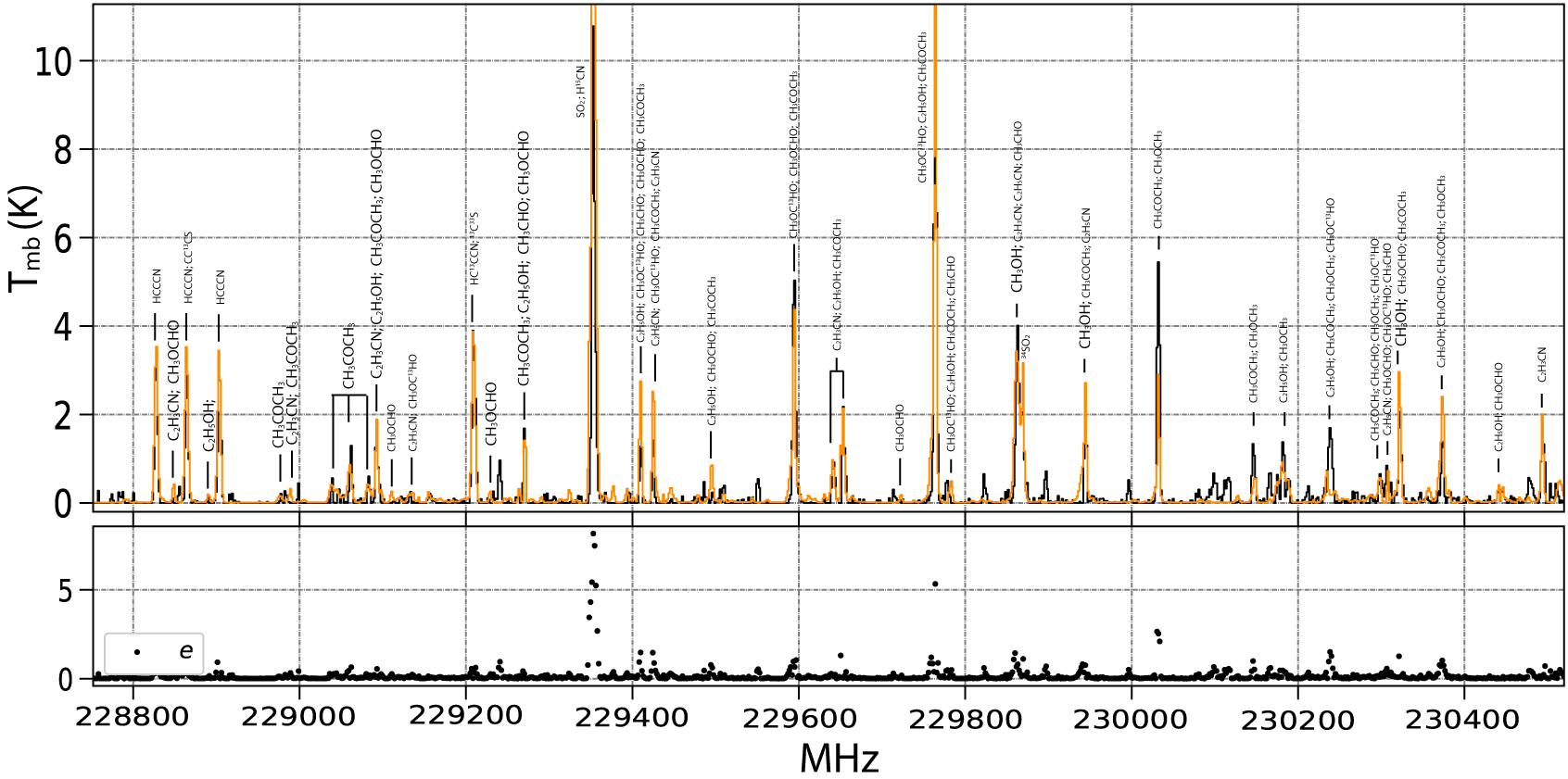}
	\includegraphics[width=\linewidth]{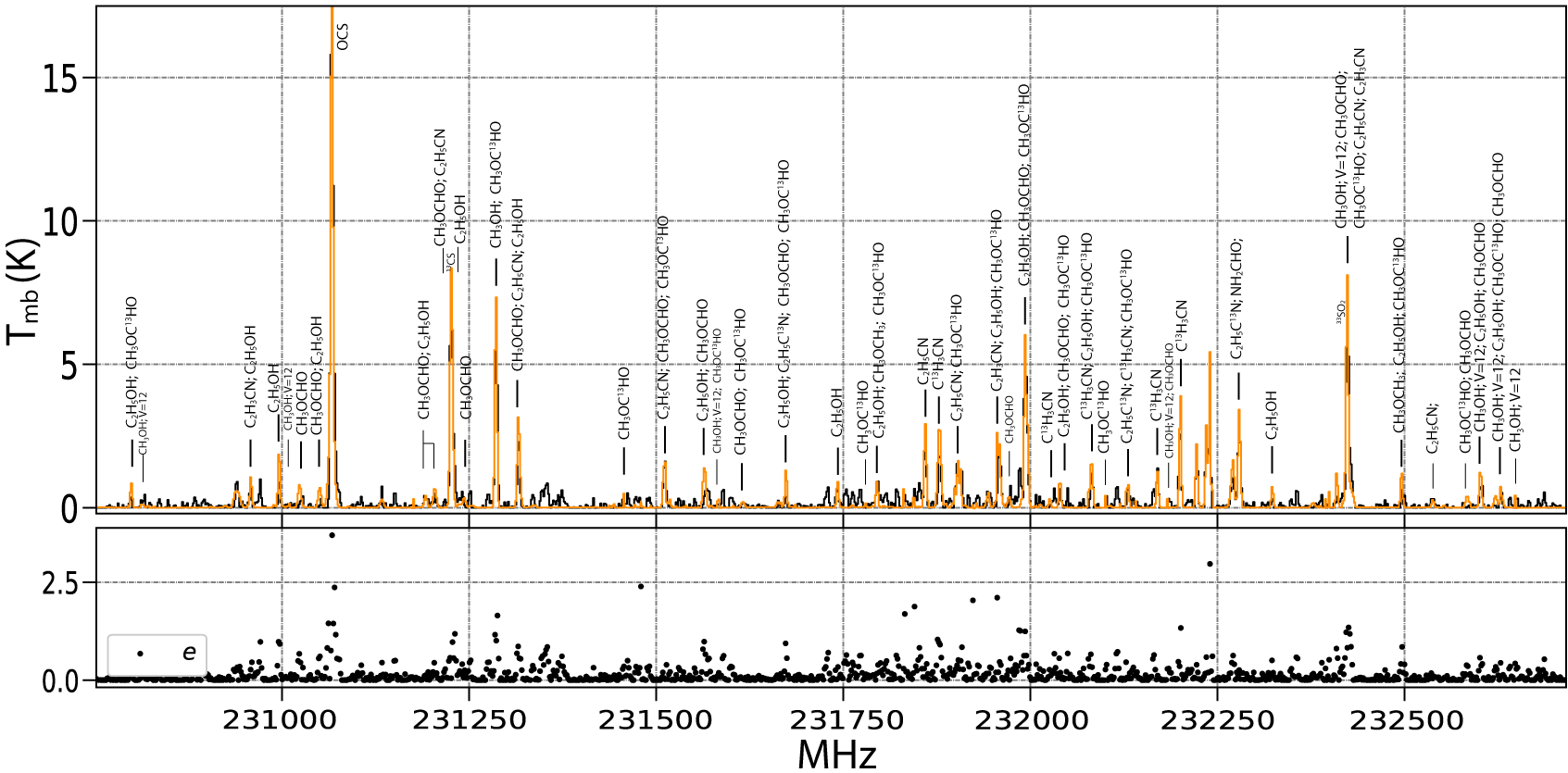}
	\caption{Continued.}
\end{figure*}

\section{Spectra of CO and SiO EHV and HV outflow detected in G351}
\begin{figure*}
\begin{center}
\includegraphics[angle=0,scale=0.6]{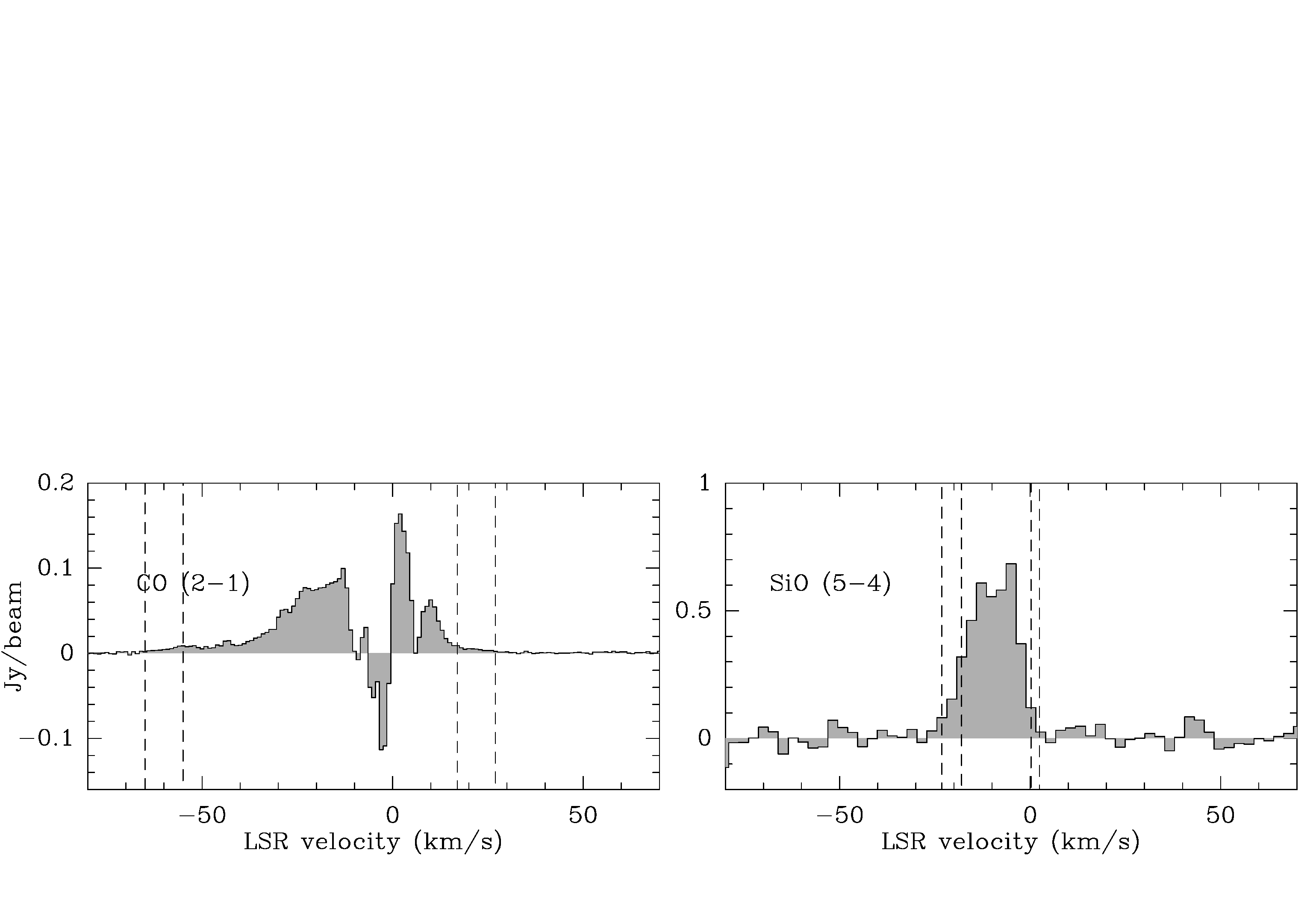}
\caption{CO (2-1) and SiO (5-4) SMA spectra in G351. The vertical dashed lines show the Extremely-High-Velocity and High-Velocity wings traced for CO and SiO, respectively. }
\label{co-sio-spect}
\end{center}
\end{figure*}
\section{Full iCOMs identified catalog}\label{APx_Full_List}
In the following we display the full set of emission lines attributed to iCOMS.

\begin{table*}
 \caption{Spectral signatures which overpass the T$_{mb}>$ 0.8 K peak emission along the Lower Sideband of the 8 GHz bandwidth coverage. For 1 to 4, the columns shows: (1) The molecular name and the corresponding vibrational excited state, (2) the rest frequency, (3) the spontaneous emission coefficient A$_{ij}$ and (4) the upper state degeneracy $g_{up}$. This organization is replicated into the columns 5 to 8. The lines are sorted by frequency.}
 \label{iCOMs_Full_over_0.8}
 \begin{tabular}{lccclccccccc}

 \toprule
		\multicolumn{8}{c}{Lower Sideband detected iCOM transitions (216 - 220 GHz)} \\
		\cmidrule(r){1-8}
\multirow{2}{1.2cm}{iCOM species}   & Frequency & A$_{ij}$ &  g$_{up}$ &\multirow{2}{1.2cm}{iCOM species}     & Frequency & A$_{ij}$&  g$_{up}$\\
    &  [GHz]& [s$^{-1}$]&   &    &  [GHz]& [s$^{-1}$]& \\
		\midrule
C$_2$H$_5$CN;v=0;	        &	216.8281  &	\num{6.9446E-07}	&	159	&	CH$_3$OCH$_3$;v=0;    	    &	219.4681  &	\num{1.2011E-05}	&	171	\\
CH$_3$OCHO;v=0;	            &	216.8302  &	\num{1.4796E-04}	&	74	&	CH$_3$OCH$_3$;v=0;    	    &	219.4681  &	\num{1.2009E-05}	&	456	\\
CH$_3$OCHO;v=0;	            &	216.8389  &	\num{1.4800E-04}	&	74	&	CH$_3$OCH$_3$;v=0;    	    &	219.4682  &	\num{1.2010E-05}	&	285	\\
CH$_3$OH;v=0;	            &	216.9456  &	\num{1.2135E-05}	&	44	&	CH$_3$COCH$_3$;v=0;	        &	219.4690  &	\num{1.3837E-04}	&	268	\\
aGg'-(CH$_2$OH)$_2$;v=0;	&	216.9466  &	\num{5.1108E-06}	&	525	&	CH$_3$COCH$_3$;v=0;	        &	219.4690  &	\num{1.3837E-04}	&	268	\\
CH$_3$OCHO;v=0;	            &	216.9630  &	\num{2.4448E-05}	&	82	&	CH$_3$COCH$_3$;v=0;	        &	219.4691  &	\num{1.3837E-04}	&	402	\\
CH$_3$OCHO;v=0;	            &	216.9642  &	\num{2.4436E-05}	&	82	&	CH$_3$COCH$_3$;v=0;	        &	219.4691  &	\num{1.3838E-04}	&	134	\\
CH$_3$OCHO;v=0;	            &	216.9648  &	\num{1.5313E-04}	&	82	&	C$_2$H$_5$CN;v=0;	        &	219.5056  &	\num{8.8747E-04}	&	49	\\
CH$_3$OCHO;v=0;	            &	216.9659  &	\num{1.5315E-04}	&	82	&	CH$_3$OH;v=0;	            &	220.0785  &	\num{2.5157E-05}	&	68	\\
CH$_3$OCHO;v=0;	            &	216.9662  &	\num{1.5313E-04}	&	82	&	CH$_3$OCHO;v=0;	            &	220.1669  &	\num{1.5235E-04}	&	70	\\
CH$_3$OCHO;v=0;	            &	216.9674  &	\num{1.5319E-04}	&	82	&	CH$_3$COCH$_3$;v=0;	        &	220.1688  &	\num{1.2724E-04}	&	390	\\
CH$_3$OCHO;v=0;	            &	216.9680  &	\num{2.4449E-05}	&	82	&	CH$_3$COCH$_3$;v=0;	        &	220.1688  &	\num{1.2722E-04}	&	650	\\
CH$_3$OCHO;v=0;	            &	216.9692  &	\num{2.4436E-05}	&	82	&	CH$_3$OCHO;v=0;	            &	220.1903  &	\num{1.5243E-04}	&	70	\\
CH$_3$OH;v12=1;	            &	217.2992  &	\num{4.2845E-05}	&	52	&	CH$_3$COCH$_3$;v=0;	        &	220.3619  &	\num{3.9256E-04}	&	720	\\
CH$_3$OH;v12=1;A	        &	217.2992  &	\num{4.2845E-05}	&	52	&	CH$_3$COCH$_3$;v=0;	        &	220.3619  &	\num{1.1580E-04}	&	720	\\
aGg'-(CH$_2$OH)$_2$;v=0;	&	217.4500  &	\num{2.5202E-04}	&	441	&	CH$_3$COCH$_3$;v=0;	        &	220.3619  &	\num{1.1580E-04}	&	720	\\
aGg'-(CH$_2$OH)$_2$;v=0;	&	217.4503  &	\num{2.5199E-04}	&	343	&	CH$_3$COCH$_3$;v=0;	        &	220.3619  &	\num{3.9256E-04}	&	720	\\
CH$_3$OC$^{13}$HO;v=0;	    &	217.6411  &	\num{1.6465E-06}	&	194	&	CH$_3$OH;v=0;	            &	220.4014  &	\num{1.1184E-05}	&	84	\\
CH$_3$OH;v12=1;	            &	217.6426  &	\num{1.8856E-05}	&	124	&	CH$_3$CN;v=0;	            &	220.4039  &	\num{4.0253E-04}	&	100	\\
CH$_3$OH;v12=1;	            &	217.6426  &	\num{1.8856E-05}	&	124	&	CH$_3$CN;v=0;	            &	220.4758  &	\num{5.1160E-04}	&	50	\\
CH$_3$OH;v12=1;A	        &	217.6426  &	\num{1.8856E-05}	&	124	&	CH$_3$CN;v=0;	            &	220.5393  &	\num{6.0805E-04}	&	50	\\
CH$_3$OH;v12=1;A	        &	217.6426  &	\num{1.8856E-05}	&	124	&	CH$_3$COCH$_3$;v=0;	        &	220.5600  &	\num{1.1974E-05}	&	688	\\
CH$_3$OC$^{13}$HO;v=0;	    &	217.6444  &	\num{1.5385E-09}	&	78	&	C$_2$H$_3$CN;v=0;	        &	220.5614  &	\num{8.8839E-04}	&	147	\\
C$_2$H$_5$CN;v=0;	        &	217.8859  &	\num{4.0390E-05}	&	97	&	CH$_3$COCH$_3$;v=0;	        &	220.5640  &	\num{4.6982E-04}	&	92	\\
CH$_3$OH;v=0;	            &	217.8865  &	\num{3.3782E-05}	&	164	&	CH$_3$COCH$_3$;v=0;	        &	220.5649  &	\num{1.8702E-06}	&	1552	\\
CH$_3$COCH$_3$;v=0;	        &	217.8882  &	\num{8.6152E-07}	&	196	&	CH$_3$COCH$_3$;v=0;	        &	220.5649  &	\num{1.8702E-06}	&	1552	\\
CH$_3$OCHO;v=0;	            &	218.2809  &	\num{1.5077E-04}	&	70	&	CH$_3$COCH$_3$;v=0;	        &	220.5687  &	\num{2.1782E-04}	&	436	\\
CH$_3$COCH$_3$;v=0;	        &	218.2814  &	\num{2.4303E-04}	&	582	&	CH$_3$C$^{13}$N;v=0;    	&	220.5704  &	\num{8.1970E-04}	&	50	\\
CH$_3$COCH$_3$;v=0;	        &	218.2814  &	\num{2.4301E-04}	&	970	&	CH$_3$CN;v=0;	            &	220.5944  &	\num{6.9180E-04}	&	100	\\
CH$_3$OCHO;v=0;	            &	218.2979  &	\num{1.5085E-04}	&	70	&	CH$_3$COCH$_3$;v=0;	        &	220.5945  &	\num{3.8023E-04}	&	1744	\\
C$_2$H$_5$CN;v=0;	        &	218.3900  &	\num{8.6655E-04}	&	49	&	CH$_3$C$^{13}$N;v=0;    	&	220.6000  &	\num{8.6486E-04}	&	50	\\
C$_2$H$_3$CN;v=0;	        &	218.3986  &	\num{7.8358E-04}	&	141	&	CH$_3$C$^{13}$N;v=0;    	&	220.6000  &	\num{8.6486E-04}	&	50	\\
C$_2$H$_3$CN;v=0;	        &	218.3986  &	\num{7.8358E-04}	&	141	&	CH$_3$COCH$_3$;v=0;	        &	220.6039  &	\num{3.1997E-04}	&	564	\\
C$_2$H$_3$CN;v=0;	        &	218.4024  &	\num{8.0490E-04}	&	141	&	CH$_3$C$^{13}$N;v=0;    	&	220.6211  &	\num{8.9719E-04}	&	50	\\
C$_2$H$_3$CN;v=0;	        &	218.4025  &	\num{8.0490E-04}	&	141	&	CH$_3$C$^{13}$N;v=0;    	&	220.6338  &	\num{9.1661E-04}	&	50	\\
CH$_3$OH;v=0;	            &	218.4400  &	\num{4.6863E-05}	&	36	&	CH$_3$C$^{13}$N;v=0;    	&	220.6381  &	\num{9.2310E-04}	&	50	\\
C$_2$H$_3$CN;v=0;	        &	218.4524  &	\num{8.2338E-04}	&	141	&	CH$_3$CN;v=0;	            &	220.6411  &	\num{7.6271E-04}	&	50	\\
C$_2$H$_3$CN;v=0;	        &	218.5881  &	\num{6.6776E-04}	&	141	&	C$_2$H$_5$CN;v=0;	        &	220.6609  &	\num{9.0141E-04}	&	51	\\
C$_2$H$_3$CN;v=0;	        &	218.5881  &	\num{6.6776E-04}	&	141	&	CH$_3$COCH$_3$;v=0;	        &	220.6617  &	\num{7.2298E-04}	&	2256	\\
CH$_3$OCHO;v=0;	            &	218.5897  &	\num{4.5602E-07}	&	102	&	CH$_3$COCH$_3$;v=0;	        &	220.6649  &	\num{4.7040E-04}	&	368	\\
CH$_3$OCHO;v=0;	            &	218.5897  &	\num{4.5602E-07}	&	102	&	CH$_3$COCH$_3$;v=0;	        &	220.6787  &	\num{9.9404E-05}	&	976	\\
CH$_3$COCH$_3$;v=0;	        &	219.2421  &	\num{3.8802E-05}	&	688	&	CH$_3$COCH$_3$;v=0;	        &	220.6787  &	\num{9.9404E-05}	&	976	\\
CH$_3$COCH$_3$;v=0;	        &	219.2421  &	\num{4.3148E-04}	&	688	&	CH$_3$CN;v=0;	            &	220.6793  &	\num{8.2096E-04}	&	50	\\
CH$_3$COCH$_3$;v=0;	        &	219.2421  &	\num{4.3148E-04}	&	688	&	C$_2$H$_3$CN;v=0;	        &	220.7071  &	\num{3.5229E-07}	&	153	\\
CH$_3$COCH$_3$;v=0;	        &	219.2421  &	\num{3.8802E-05}	&	688	&	CH$_3$CN;v=0;	            &	220.7090  &	\num{8.6614E-04}	&	100	\\
C$_2$H$_3$CN;v=0;	        &	219.4006  &	\num{8.6068E-04}	&	141	&	CH$_3$OCHO;v=0;	            &	220.7105  &	\num{2.0885E-06}	&	98	\\
CH$_3$OCH$_3$;v=0;    	    &	219.4681  &	\num{1.2011E-05}	&	114	&	CH$_3$OCHO;v=0;	            &	220.7105  &	\num{3.5248E-06}	&	98	\\
\bottomrule 
 \end{tabular}
\end{table*}

\begin{table*}
 \contcaption{ Spectral signatures as before, but for the Upper Sideband transitions.}
 \begin{tabular}{lccclccccccc}

 \toprule

		\multicolumn{8}{c}{Upper Sideband detected iCOM transitions (228 - 232 GHz)} \\
		\cmidrule(r){1-8}
\multirow{2}{1.2cm}{iCOM species}   & Frequency & A$_{ij}$ &  g$_{up}$ &\multirow{2}{1.2cm}{iCOM species}     & Frequency & A$_{ij}$&  g$_{up}$\\
    &  [GHz]& [s$^{-1}$]&   &    &  [GHz]& [s$^{-1}$]& \\
		\midrule
CH$_3$OH;v=0;	            &	232.4186    &	\num{1.8675E-05}	&	84	&	C$_2$H$_5$OH;v=0;    	        &	231.6687    &	\num{1.1080E-04}	&	29	\\
CH$_3$OH;v=0;	            &	231.2811    &	\num{1.8314E-05}	&	84	&	C$_2$H$_5$OH;v=0;    	        &	230.9538    &	\num{7.7649E-05}	&	33	\\
CH$_3$OCHO;v=0;	            &	231.0190    &	\num{1.0327E-05}	&	50	&	C$_2$H$_5$OH;v=0;    	        &	232.5966    &	\num{8.0717E-05}	&	29	\\
CH$_3$OCHO;v=0;	            &	232.5973    &	\num{1.5301E-05}	&	78	&	C$_2$H$_5$OH;v=0;    	        &	232.4048    &	\num{8.0221E-05}	&	35	\\
CH$_3$OCHO;v=0;	            &	231.9552    &	\num{1.5602E-05}	&	82	&	C$_2$H$_5$OH;v=0;    	        &	232.0346    &	\num{8.1032E-05}	&	37	\\
CH$_3$OCHO;v=0;	            &	231.5613    &	\num{4.0319E-06}	&	110	&	C$_2$H$_5$OH;v=0;    	        &	231.7376    &	\num{8.1839E-05}	&	39	\\
CH$_3$OCHO;v=0;	            &	231.5597    &	\num{7.2245E-06}	&	110	&	C$_2$H$_5$OH;v=0;    	        &	231.5609    &	\num{8.2653E-05}	&	41	\\
CH$_3$OH;v12=1;    	        &	232.4228    &	\num{5.5499E-06}	&	15	&	C$_2$H$_5$OH;v=0;    	        &	231.5586    &	\num{8.3570E-05}	&	43	\\
CH$_3$OH;v12=1;    	        &	232.6248    &	\num{6.5434E-06}	&	31	&	C$_2$H$_5$OH;v=0;    	        &	231.7901    &	\num{8.4615E-05}	&	45	\\
CH$_3$OCH$_3$;v=0;    	    &	231.9877    &	\num{7.2397E-05}	&	135	&	C$_2$H$_5$OH;v=0;    	        &	231.3124    &	\num{1.6987E-07}	&	55	\\
CH$_3$OCH$_3$;v=0;    	    &	231.9878    &	\num{7.2394E-05}	&	216	&	C$_2$H$_5$OH;v=0;    	        &	231.7898    &	\num{2.2387E-05}	&	63	\\
CH$_3$OCH$_3$;v=0;    	    &	231.9879    &	\num{7.2389E-05}	&	81	&	C$_2$H$_5$OH;v=0;    	        &	231.8408    &	\num{1.0785E-05}	&	71	\\
CH$_3$OCH$_3$;v=0;    	    &	231.9879    &	\num{7.2404E-05}	&	54	&	C$_2$H$_5$OH;v=0;    	        &	231.9519    &	\num{2.0151E-06}	&	67	\\
C$_2$H$_3$CN;v=0;	        &	230.9515    &	\num{2.9486E-05}	&	117	&	C$_2$H$_5$OH;v=0;    	        &	231.9519    &	\num{2.0151E-06}	&	67	\\
C$_2$H$_3$CN;v=0;	        &	230.7386    &	\num{1.0185E-03}	&	153	&	C$_2$H$_5$OH;v=0;    	        &	231.9925    &	\num{1.0037E-05}	&	93	\\
C$_2$H$_3$CN;v=0;	        &	231.9523    &	\num{1.0284E-03}	&	147	&	C$_2$H$_5$OH;v=0;    	        &	231.9925    &	\num{1.0037E-05}	&	93	\\
C$_2$H$_3$CN;v=0;	        &	232.4165    &	\num{7.8130E-08}	&	225	&	C$_2$H$_5$C$^{13}$N;v=0;	        &	232.2723    &	\num{1.0033E-03}	&	53	\\
C$_2$H$_5$CN;v=0;	        &	231.3132    &	\num{5.4121E-05}	&	49	&	C$_2$H$_5$C$^{13}$N;v=0;	        &	231.9885    &	\num{6.9264E-04}	&	53	\\
C$_2$H$_5$CN;v=0;	        &	231.3104    &	\num{1.0413E-03}	&	53	&	C$_2$H$_5$C$^{13}$N;v=0;	        &	231.9885    &	\num{6.9264E-04}	&	53	\\
C$_2$H$_5$CN;v=0;	        &	231.9904    &	\num{1.0550E-03}	&	55	&	NH$_2$CHO;v=0;	            &	232.2743    &	\num{8.8168E-04}	&	138	\\
C$_2$H$_5$CN;v=0;	        &	232.2154    &	\num{1.5268E-06}	&	51	&	CH$_3$OH;v12=1;    	    &	229.8665    &	\num{2.0083E-05}	&	77	\\
C$_2$H$_5$CN;v=0;	        &	231.3123    &	\num{9.1831E-05}	&	55	&	CH$_3$OH;v12=1;A	    &	229.8665    &	\num{1.9898E-05}	&	308	\\
C$_2$H$_5$CN;v=0;	        &	231.8542    &	\num{1.0530E-03}	&	55	&	CH$_3$OCHO;v=0;	        &	229.4203    &	\num{1.7520E-04}	&	74	\\
C$_2$H$_5$CN;v=0;	        &	232.4206    &	\num{7.2673E-05}	&	125	&	CH$_3$OCHO;v=0;	        &	229.4050    &	\num{1.7515E-04}	&	74	\\
CH$_3$OC$^{13}$HO;v=0;	    &	232.4178    &	\num{8.4218E-07}	&	38	&	CH$_3$OCHO;v=0;	        &	229.5905    &	\num{1.7779E-05}	&	78	\\
CH$_3$OC$^{13}$HO;v=0;	    &	231.9870    &	\num{6.9214E-06}	&	62	&	CH$_3$OCHO;v=0;	        &	229.4897    &	\num{7.8740E-07}	&	134	\\
CH$_3$OC$^{13}$HO;v=0;	    &	231.9871    &	\num{6.9215E-06}	&	62	&	CH$_3$CHO;v=0;	            &	229.8609    &	\num{4.0766E-05}	&	42	\\
CH$_3$OC$^{13}$HO;v=0;	    &	231.6690    &	\num{7.2844E-06}	&	66	&	CH$_3$CHO;v=0;	            &	230.3019    &	\num{4.1938E-04}	&	50	\\
CH$_3$OC$^{13}$HO;v=0;	    &	231.6695    &	\num{7.2845E-06}	&	66	&	CH$_3$CHO;v=0;	            &	230.3158    &	\num{4.1933E-04}	&	50	\\
CH$_3$OC$^{13}$HO;v=0;	    &	231.2841    &	\num{7.5869E-06}	&	70	&	CH$_3$CHO;v=0;	            &	229.4181    &	\num{1.1033E-05}	&	154	\\
CH$_3$OC$^{13}$HO;v=0;	    &	232.4051    &	\num{6.9938E-05}	&	78	&	CH$_3$CHO;v=0;	            &	229.4176    &	\num{1.1033E-05}	&	154	\\
CH$_3$OC$^{13}$HO;v=0;	    &	231.9197    &	\num{5.7783E-05}	&	78	&	C$_2$H$_3$CN;v=0;	    &	230.4879    &	\num{1.0131E-03}	&	147	\\
CH$_3$OC$^{13}$HO;v=0;	    &	231.4764    &	\num{3.9773E-08}	&	98	&	C$_2$H$_3$CN;v=0;	    &	229.6478    &	\num{1.0037E-03}	&	153	\\
CH$_3$OC$^{13}$HO;v=0;	    &	231.9928    &	\num{3.1142E-08}	&	70	&	C$_2$H$_3$CN;v=0;	    &	229.0870    &	\num{9.8193E-04}	&	147	\\
CH$_3$OC$^{13}$HO;v=0;	    &	231.9549    &	\num{3.1127E-08}	&	70	&	CH$_3$OC$^{13}$HO;v=0;	&	229.5921    &	\num{8.2046E-06}	&	82	\\
CH$_3$OC$^{13}$HO;v=0;	    &	232.2366    &	\num{3.5009E-06}	&	150	&	CH$_3$OC$^{13}$HO;v=0;	&	229.7575    &	\num{2.6452E-06}	&	118	\\
CH$_3$OC$^{13}$HO;v=0;	    &	230.7416    &	\num{9.4075E-06}	&	162	&	C$_2$H$_5$CN;v=0;	    &	229.2652    &	\num{1.0122E-03}	&	53	\\
CH$_3$OC$^{13}$HO;v=0;	    &	232.4913    &	\num{2.0340E-06}	&	170	&	C$_2$H$_5$CN;v=0;	    &	229.4198    &	\num{3.4398E-05}	&	87	\\
C$_2$H$_5$CN-15;v=0;    	&	231.8404    &	\num{6.4916E-05}	&	17	&	C$_2$H$_5$CN;v=0;	    &	229.7519    &	\num{5.7938E-05}	&	97	\\
C$_2$H$_5$CN-15;v=0;    	&	231.8274    &	\num{6.4913E-05}	&	17	&	C$_2$H$_5$CN;v=0;	    &	229.4042    &	\num{4.6438E-06}	&	159	\\
C$_2$H$_5$CN-15;v=0;    	&	231.4751    &	\num{4.0946E-05}	&	27	&	C$_2$H$_5$OH;v=0;    	        &	230.6726    &	\num{1.0602E-04}	&	27	\\
C$_2$H$_5$CN-15;v=0;    	&	231.9190    &	\num{4.1796E-05}	&	29	&	C$_2$H$_5$OH;v=0;    	        &	229.4911    &	\num{7.7558E-05}	&	35	\\
C$^{13}$H$_3$CN;v=0;	            &	232.2342    &	\num{1.0796E-03}	&	54	&	C$_2$H$_5$OH;v=0;    	        &	229.6494    &	\num{6.5603E-07}	&	29	\\
C$^{13}$H$_3$CN;v=0;	            &	232.2298    &	\num{1.0733E-03}	&	54	&	C$_2$H$_5$OH;v=0;    	        &	229.0907    &	\num{2.5753E-06}	&	49	\\
C$^{13}$H$_3$CN;v=0;	            &	232.2167    &	\num{1.0539E-03}	&	54	&	C$_2$H$_5$OH;v=0;    	        &	229.0888    &	\num{3.4598E-07}	&	71	\\
C$^{13}$H$_3$CN;v=0;	            &	232.1949    &	\num{1.0217E-03}	&	54	&	C$_2$H$_5$OH;v=0;    	        &	229.0892    &	\num{6.1540E-08}	&	73	\\
C$^{13}$H$_3$CN;v=0;	            &	232.1949    &	\num{1.0217E-03}	&	54	&	C$_2$H$_5$OH;v=0;    	        &	230.1766    &	\num{2.0084E-07}	&	91	\\
C$^{13}$H$_3$CN;v=0;	            &	232.1644    &	\num{9.7649E-04}	&	54	&	C$_2$H$_5$OH;v=0;    	        &	230.1780    &	\num{5.1910E-06}	&	91	\\
C$^{13}$H$_3$CN;v=0;	            &	232.1251    &	\num{9.1870E-04}	&	54	&	C$_2$H$_5$OH;v=0;    	        &	229.4225    &	\num{1.0946E-04}	&	91	\\
C$^{13}$H$_3$CN;v=0;	            &	232.0772    &	\num{8.4805E-04}	&	54	&	C$_2$H$_5$OH;v=0;    	        &	229.7549    &	\num{3.7369E-07}	&	89	\\
C$^{13}$H$_3$CN;v=0;	            &	232.0772    &	\num{8.4805E-04}	&	54	&	C$_2$H$_5$OH;v=0;    	        &	229.4915    &	\num{4.6090E-06}	&	93	\\
C$^{13}$H$_3$CN;v=0;	            &	231.9554    &	\num{6.6828E-04}	&	54	&	C$_2$H$_5$OH;v=0;    	        &	229.0886    &	\num{1.3346E-05}	&	73	\\
C$^{13}$H$_3$CN;v=0;A	            &	232.2342    &	\num{1.0796E-03}	&	54	&	C$_2$H$_5$OH;v=0;    	        &	229.0886    &	\num{1.3346E-05}	&	73	\\
C$^{13}$H$_3$CN;v=0;A	            &	232.1949    &	\num{1.0217E-03}	&	54	&	CH$_3$OCH$_3$;v=0;    	&	230.3679    &	\num{1.4057E-05}	&	57	\\
C$^{13}$H$_3$CN;v=0;A	            &	232.1949    &	\num{1.0217E-03}	&	54	&	CH$_3$OCH$_3$;v=0;    	&	230.3679    &	\num{1.4056E-05}	&	114	\\
C$^{13}$H$_3$CN;v=0;A	            &	232.0772    &	\num{8.4805E-04}	&	54	&	CH$_3$OCH$_3$;v=0;    	&	230.3682    &	\num{1.4057E-05}	&	456	\\
C$^{13}$H$_3$CN;v=0;A	            &	232.0772    &	\num{8.4805E-04}	&	54	&	CH$_3$OCH$_3$;v=0;    	&	230.3685    &	\num{1.4056E-05}	&	171	\\
\bottomrule 
 \end{tabular}
\end{table*}

\begin{table*}
 \contcaption{ Spectral signatures for the Upper Sideband transitions.}
 \begin{tabular}{lccclccccccc}

 \toprule

		\multicolumn{8}{c}{Upper Sideband detected iCOM transitions (228 - 232 GHz)} \\
		\cmidrule(r){1-8}
\multirow{2}{1.2cm}{iCOM species}   & Frequency & A$_{ij}$ &  g$_{up}$ &\multirow{2}{1.2cm}{iCOM species}     & Frequency & A$_{ij}$&  g$_{up}$\\
    &  [GHz]& [s$^{-1}$]&   &    &  [GHz]& [s$^{-1}$]& \\
		\midrule
C$^{13}$H$_3$CN;v=0;E	            &	232.2298    &	\num{1.0733E-03}	&	54	&	CH$_3$OCH$_3$;v=0;    	&	230.0276    &	\num{1.1771E-05}	&	178	\\
C$^{13}$H$_3$CN;v=0;E	            &	232.2167    &	\num{1.0539E-03}	&	54	&	CH$_3$COCH$_3$;v=0;	    &	229.6500    &	\num{3.5563E-04}	&	150	\\
C$^{13}$H$_3$CN;v=0;E	            &	232.1644    &	\num{9.7649E-04}	&	54	&	CH$_3$COCH$_3$;v=0;	    &	229.0558    &	\num{4.4249E-04}	&	720	\\
C$^{13}$H$_3$CN;v=0;E	            &	232.1251    &	\num{9.1870E-04}	&	54	&	CH$_3$COCH$_3$;v=0;	    &	229.0558    &	\num{9.6583E-05}	&	720	\\
C$^{13}$H$_3$CN;v=0;E	            &	231.9554    &	\num{6.6828E-04}	&	54	&	CH$_3$COCH$_3$;v=0;	    &	229.0558    &	\num{8.2073E-05}	&	720	\\
C$_2$H$_3$CN;v=0;    	    &	230.9528    &	\num{7.2866E-08}	&	41	&	CH$_3$COCH$_3$;v=0;	    &	229.0558    &	\num{4.5698E-04}	&	720	\\
C$_2$H$_3$CN;v=0;    	    &	230.9528    &	\num{8.0744E-08}	&	37	&	CH$_3$COCH$_3$;v=0;	    &	230.1767    &	\num{1.2709E-04}	&	752	\\
C$_2$H$_3$CN;v=0;    	    &	230.7385    &	\num{1.0051E-03}	&	51	&	CH$_3$COCH$_3$;v=0;	    &	230.1767    &	\num{4.5352E-04}	&	752	\\
C$_2$H$_3$CN;v=0;    	    &	230.7401    &	\num{1.6764E-06}	&	49	&	CH$_3$COCH$_3$;v=0;	    &	230.1767    &	\num{4.5352E-04}	&	752	\\
C$_2$H$_3$CN;v=0;    	    &	230.7370    &	\num{1.6106E-06}	&	51	&	CH$_3$COCH$_3$;v=0;	    &	230.1767    &	\num{1.2709E-04}	&	752	\\
C$_2$H$_3$CN;v=0;    	    &	230.7385    &	\num{1.0050E-03}	&	49	&	CH$_3$COCH$_3$;v=0;	    &	230.1784    &	\num{1.8896E-06}	&	78	\\
C$_2$H$_3$CN;v=0;    	    &	230.7386    &	\num{1.0067E-03}	&	53	&	CH$_3$COCH$_3$;v=0;	    &	230.0284    &	\num{1.1370E-06}	&	234	\\
C$_2$H$_3$CN;v=0;    	    &	231.9523    &	\num{1.0148E-03}	&	49	&	CH$_3$COCH$_3$;v=0;	    &	229.5900    &	\num{2.0900E-06}	&	172	\\
C$_2$H$_3$CN;v=0;    	    &	231.9537    &	\num{1.8399E-06}	&	47	&	CH$_3$COCH$_3$;v=0;	    &	230.3686    &	\num{1.2738E-06}	&	752	\\
C$_2$H$_3$CN;v=0;    	    &	231.9510    &	\num{1.7647E-06}	&	49	&	CH$_3$COCH$_3$;v=0;	    &	229.4188    &	\num{2.2872E-06}	&	750	\\
C$_2$H$_3$CN;v=0;    	    &	231.9523    &	\num{1.0146E-03}	&	47	&	CH$_3$COCH$_3$;v=0;	    &	230.3690    &	\num{1.1679E-04}	&	364	\\
C$_2$H$_3$CN;v=0;    	    &	231.9523    &	\num{1.0165E-03}	&	51	&	CH$_3$COCH$_3$;v=0;	    &	229.2671    &	\num{1.3434E-04}	&	372	\\
C$_2$H$_3$CN;v=0;    	    &	232.4166    &	\num{7.6814E-08}	&	75	&	CH$_3$COCH$_3$;v=0;	    &	229.7553    &	\num{2.9944E-04}	&	1616	\\
C$_2$H$_3$CN;v=0;    	    &	232.4165    &	\num{7.6856E-08}	&	73	&	CH$_3$COCH$_3$;v=0;	    &	230.1767    &	\num{2.9404E-04}	&	762	\\
C$_2$H$_3$CN;v=0;    	    &	232.4165    &	\num{7.6809E-08}	&	77	&	CH$_3$COCH$_3$;v=0;	    &	229.7530    &	\num{3.4562E-04}	&	540	\\
C$_2$H$_5$OH;v=0;    	            &	230.9914    &	\num{1.1962E-04}	&	29	&	CH$_3$COCH$_3$;v=0;	    &	229.7547    &	\num{3.4564E-04}	&	270	\\
C$_2$H$_5$OH;v=0;    	            &	230.7939    &	\num{6.2028E-05}	&	13	&	CH$_3$OH;v=0;	        &	229.7588    &	\num{4.1912E-05}	&	68	\\
C$_2$H$_5$OH;v=0;    	            &	230.7938    &	\num{6.2028E-05}	&	13	&	CH$_3$OH;v=0;	        &	229.5891    &	\num{2.0839E-05}	&	124	\\
C$_2$H$_5$OH;v=0;    	            &	230.9929    &	\num{3.9892E-07}	&	13	&	CH$_3$OH;v=0;	        &	229.9392    &	\num{2.0675E-05}	&	156	\\
C$_2$H$_5$OH;v=0;    	            &	230.9924    &	\num{3.9892E-07}	&	13	&	CH$_3$OH;v=0;	        &	229.8642    &	\num{2.0650E-05}	&	156	\\
C$_2$H$_5$OH;v=0;    	            &	232.0758    &	\num{7.7291E-05}	&	31	&	CH$_3$OH;v=0;	        &	230.3687    &	\num{2.0794E-05}	&	180	\\
C$_2$H$_5$OH;v=0;    	            &	232.4913    &	\num{1.1210E-04}	&	29	&		&				&		&		\\
\bottomrule 
 \end{tabular}
\end{table*}

\bsp	
\label{lastpage}
\end{document}